\documentclass[aps,prb,preprint,showpacs,showkeys,floatfix]{revtex4}

\usepackage{graphicx,color}
\usepackage{multirow,slashbox}

\graphicspath{{figs/}}
\begin{document}
	

\title{Intrinsic characteristic radius drives phonon anomalies in Janus transition metal dichalcogenide nanotubes}



\author{Jing-Jing Zhang}
\affiliation{School of Mechatronic Engineering and Automation, Shanghai University, Shanghai 200444, People's Republic of China}

\author{Jin-Wu Jiang}
\altaffiliation{Corresponding author: jiangjinwu@shu.edu.cn; jwjiang5918@hotmail.com}
\affiliation{Shanghai Key Laboratory of Mechanics in Energy Engineering, Shanghai Institute of Applied Mathematics and Mechanics, Shanghai Frontier Science Center of Mechanoinformatics, School of Mechanics and Engineering Science, Shanghai University, Shanghai 200072, People’s Republic of China}

\date{\today}
\begin{abstract}

Transition metal dichalcogenides and their derivatives offer a versatile platform for exploring novel structural and functional properties in low-dimensional materials. In particular, Janus monolayers possess an intrinsic out-of-plane asymmetry that induces a built-in bending radius, which can strongly influence their physical behavior. In this work, we investigate the tubular structures formed by rolling Janus monolayers into the Janus nanotube with an extrinsic radius. Using a combination of atomistic simulations and continuum mechanics, we identify that the total energy of the Janus nanotube is minimized when the tube radius equals to the intrinsic bending radius of the Janus monolayer. An analytical expression for this characteristic radius is derived, providing a theoretical basis for understanding the stability of Janus nanotubes. Furthermore, we find that the optical phonon modes in these Janus nanotubes exhibit an anomalous dependence on the tube radius; i.e., their frequencies reach a maximum value near the characteristic radius, in contrast to the monotonic increase of optical phonon frequencies with radius in conventional nanotubes. The phonon anomaly is due to the soft phonon mode effect induced by the deviation from the most stable tubular configuration with the characteristic radius. These results uncover a unique coupling between intrinsic and extrinsic curvature in Janus systems and open new pathways for tuning vibrational and other properties in curved low-dimensional materials.

\end{abstract}
\maketitle
\pagebreak

\section{Introduction}

Transition metal dichalcogenides (TMDs), such as molybdenum disulfide (MoS$_2$), have attracted tremendous attention owing to their unique structural, electronic, optical, and mechanical properties.\cite{manzeli20172d,choi2017recent,wang2024critical,gupta2025two} By selectively substituting different transition metals or chalcogen elements, a rich family of heterostructures can be obtained, including out-of-plane van der Waals heterojunctions,\cite{han2018van,ma2024van} in-plane heterostructures,\cite{DuanX2014nn,wang2019recent,liu20252d} and Janus structures with broken out-of-plane symmetry.\cite{cheng2013spin,lu2017janus,zhang2017janus,zhu2025robust} These structural modifications provide a valuable platform for tuning the intrinsic properties of TMDs, making them highly promising for applications in next-generation electronic and optoelectronic devices.

Among these structures, Janus monolayers are particularly interesting because they exhibit an inherent built-in curvature that originates from their structural asymmetry.\cite{xiong2018spontaneous,ye2020intrinsic,yang2024unveiling} This intrinsic curvature plays a critical role in modulating various material properties, such as piezoelectricity,\cite{dong2017large,rawat2020nanoscale} tribo-piezoelectricity,\cite{cai2019tribo,li2022electro} spin–orbit coupling,\cite{smaili2021janus,yu2021spin,lin2025rashba,li2025key} and catalytic activity.\cite{er2018prediction} Understanding and harnessing this curvature effect has therefore become a central theme in the study of Janus TMDs.

Tubular architectures, on the other hand, are often more stable than their two-dimensional planar counterparts and provide additional degrees of freedom for tailoring material behavior.\cite{popov2004carbon,popov2006radius,XiangR2020sci,WangP2020acsn,CambreS2021small} The tube radius can be continuously tuned, enabling systematic control over electronic band structures, strain fields, and mechanical responses. As such, tubular nanostructures have been widely explored for both fundamental studies and potential applications.\cite{tang2024chirality}

When a Janus monolayer is rolled into a tubular geometry, its intrinsic curvature becomes coupled with the extrinsic curvature imposed by the tube. This interplay may give rise to entirely new physical phenomena that are absent in either the Janus monolayer or symmetric nanotubes. Exploring this coupling between intrinsic and extrinsic curvature thus opens an exciting pathway toward discovering novel functionalities in low-dimensional materials.

In this work, we focus on the tubular structures formed by rolling Janus monolayers, hereafter referred to as Janus nanotubes. We demonstrate that there exists a characteristic radius at which the nanotube attains the lowest total energy, thereby representing the most stable configuration. A theoretical framework based on continuum mechanics is developed to analytically derive this characteristic radius and elucidate its physical origin. Furthermore, we reveal that this characteristic radius plays a decisive role in determining the vibrational properties of Janus nanotubes. More specifically, the optical phonon modes exhibit an anomalous dependence on tube radius, showing a maximum frequency near the characteristic radius, resulting from the soft mode effect induced by the deviation from the most stable tubular configuration with characteristic radius. This behavior stands in sharp contrast to conventional nanotubes, where the optical phonon frequencies generally increase monotonically with radius.

\section{Structure and simulation details}

\begin{figure}[tb]
	\begin{center}
		\scalebox{1.5}[1.5]{\includegraphics[width=8cm]{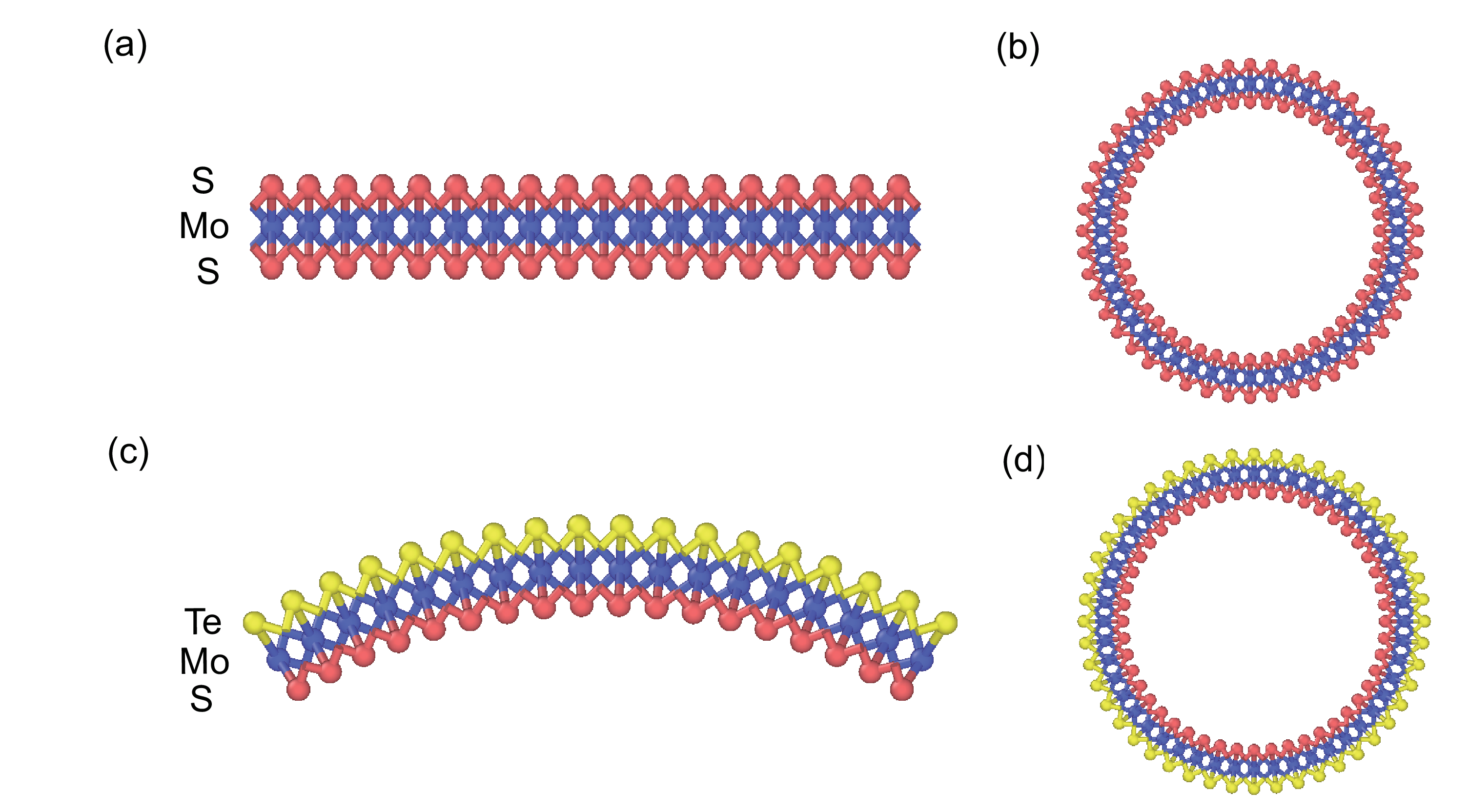}}
	\end{center}
	\caption{(Color online) Representative structures studied in this work. (a) The flat MoS$_2$ monolayer. (b) The MoS$_2$ nanotube with radius 12.1~{\AA}. (c) The MoSTe Janus monolayer with intrinsic bending radius of $R_C = 25.8$~{\AA}. (d) The MoSTe Janus nanotube with radius 12.8~{\AA}.}
	\label{fig_structure}
\end{figure}

Major TMD structures studied in this work are illustrated in Fig.~\ref{fig_structure}. Fig.~\ref{fig_structure}~(a) is the side view of the MoS$_2$ monolayer, which is flat after energy minimization due to the symmetric atomic configuration. Fig.~\ref{fig_structure}~(b) shows the top view of the MoS$_2$ nanotube with radius $r=12.1$~{\AA}. Fig.~\ref{fig_structure}~(c) shows the intrinsic bending configuration of the Janus MoSTe monolayer, with an intrinsic bending radius $R_C=25.8$~{\AA}. Fig.~\ref{fig_structure}~(d) shows the top view of the Janus MoSTe nanotube with radius $r=12.8$~{\AA}. Note that the present work studies a series of similar TMD structures with formula MX$_2$ (M=Mo, W; X=S, Se, Te), though we have only displayed MoS$_2$ and MoSTe in Fig.~\ref{fig_structure}.

Numerical simulations are carried out using the Large-scale Atomic Molecular Massively Parallel Simulator (LAMMPS),\cite{PlimptonSJ} and atomic trajectories were visualized and analyzed with the OVITO package.\cite{ovito} The interatomic interactions are modeled using the Stillinger–Weber potential.\cite{StillingerFH,JiangJW2018swmx2} The structure is relaxed by the energy minimization with the conjugate gradient algorithm.\cite{PerdewJP1996prl} The phonon dispersion is calculated with the GULP package.\cite{gulp}

\section{Intrinsic characteristic radius}

We now examine the intrinsic bending phenomenon observed in Fig.~\ref{fig_structure}~(c). We will present numerical simulation results to clearly display this effect and then discuss the underlying mechanism.

\subsection{Numerical simulation results}

\begin{figure}[tb]
	\begin{center}
		\scalebox{1}[1]{\includegraphics[width=8cm]{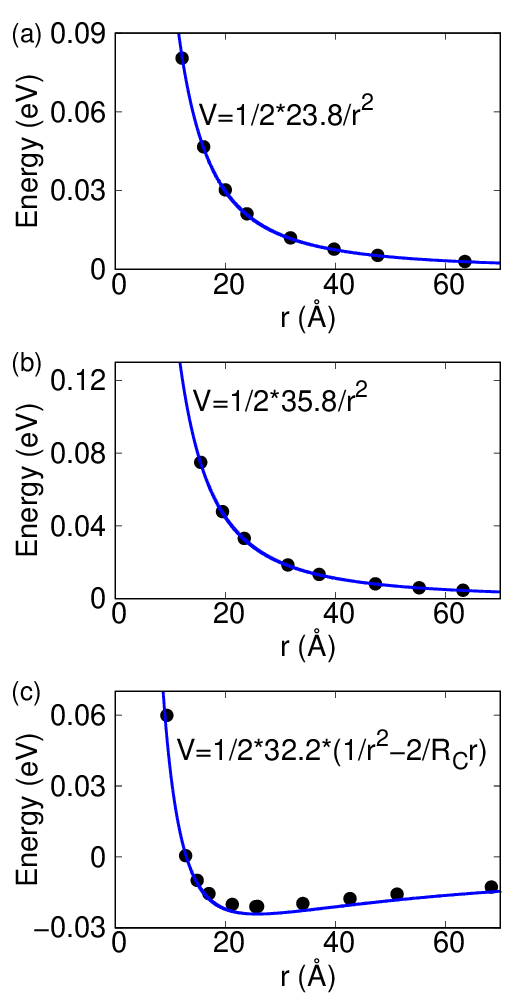}}
	\end{center}
	\caption{(Color online) Radial dependence for the energy per atom of the TMD nanotubes with respective to the monolayer planar structure for (a) MoS$_2$, (b) MoTe$_2$, and (c) MoSTe.}
	\label{fig_energy}
\end{figure}

As shown in Fig.~\ref{fig_structure}, the Janus MoSTe monolayer will be bent spontaneously after energy minimization process. The spontaneous bending radius for the Janus MoSTe monolayer is $R_C=25.8$~{\AA}. Similar spontaneous bending phenomenon has been obtained by several previous works.\cite{xiong2018spontaneous,ye2020intrinsic,yang2024unveiling}

The spontaneous bending phenomenon indicates that the MoSTe monolayer will be most stable if the monolayer is bent properly. It can be imagine that the MoSTe nanotube shall have the lowest energy if the nanotube's radius equals to the spontaneous bending radius of the MoSTe monolayer. We thus calculate the difference between the potential energy per atom of the tubular structure and the planar monolayer, which is essentially the bending energy, as shown in Fig.~\ref{fig_energy}. For pure structures MoS$_2$ and MoTe$_2$, the bending energy decreases with increasing radius as a parabolic function,\cite{arroyo2004finite,jiang2013elastic}
\begin{eqnarray}
	V_{B}^{\rm pure} & = & \frac{1}{2}D \frac{1}{r^2},
	\label{eq_vb_pure}
\end{eqnarray}
where $D$ is the bending stiffness for the pure TMD monolayer and $r$ is the radius for the nanotube. Figs.~\ref{fig_energy} (a) and (b) show that the bending energy in the pure structures MoS$_2$ and MoTe$_2$ can be well described by Eq.~(\ref{eq_vb_pure}).

For the Janus MoSTe nanotube, the radial dependence for the potential energy is shown in Fig.~\ref{fig_energy}~(c). It is obvious that there is a minima in the potential energy at $r=R_C=25.8$~{\AA}, which is exactly the value of the intrinsic bending radius of the Janus MoSTe monolayer. Hence, the spontaneous bending radius $R_C$ can be regarded as a characteristic radius of the Janus MoSTe nanotube, at which the nanotube is most stable. The potential energy increases when the nanotube's radius deviates from $R_C$. In the limit of large radius, the potential energy approaches to the value of planar MoSTe monolayer. We find that the potential energy can be well fitted by the following formula
\begin{eqnarray}
	V_{B}^{\rm janus} & = & \frac{1}{2}D (\frac{1}{r^2}-\frac{2}{R_C r}),
	\label{eq_vb_janus}
\end{eqnarray}
where $D$ is the effective bending stiffness for the TMD Janus monolayer, and $R_C$ is the characteristic radius.

\subsection{Analytic explanation}

In the above, we have demonstrated through numerical simulations that there is a characteristic radius for the Janus nanotube, at which the structure has the lowest energy. We now derive the analytic formula for the characteristic radius.

The lattice constant for MoTe$_2$ is larger than MoS$_2$ as the Te atom is larger than the S atom. As a result, if the MoSTe Janus structure keeps in the planar structure, the Te atomic layer will be compressed, while the S atomic layer will be stretched. To relax the strain energy stored within Te and S atomic layers, the MoSTe Janus monolayer will be bent spontaneously, with the larger Te atom on the outer surface while the smaller S atom on the inner surface. Based on this mechanism, we now provide an analytic model to explain the spontaneous bending radius of the Janus MoSTe monolayer. There are two steps for the spontaneous bending phenomenon to occur.

\begin{figure}[tb]
	\begin{center}
		\scalebox{1.5}[1.5]{\includegraphics[width=8cm]{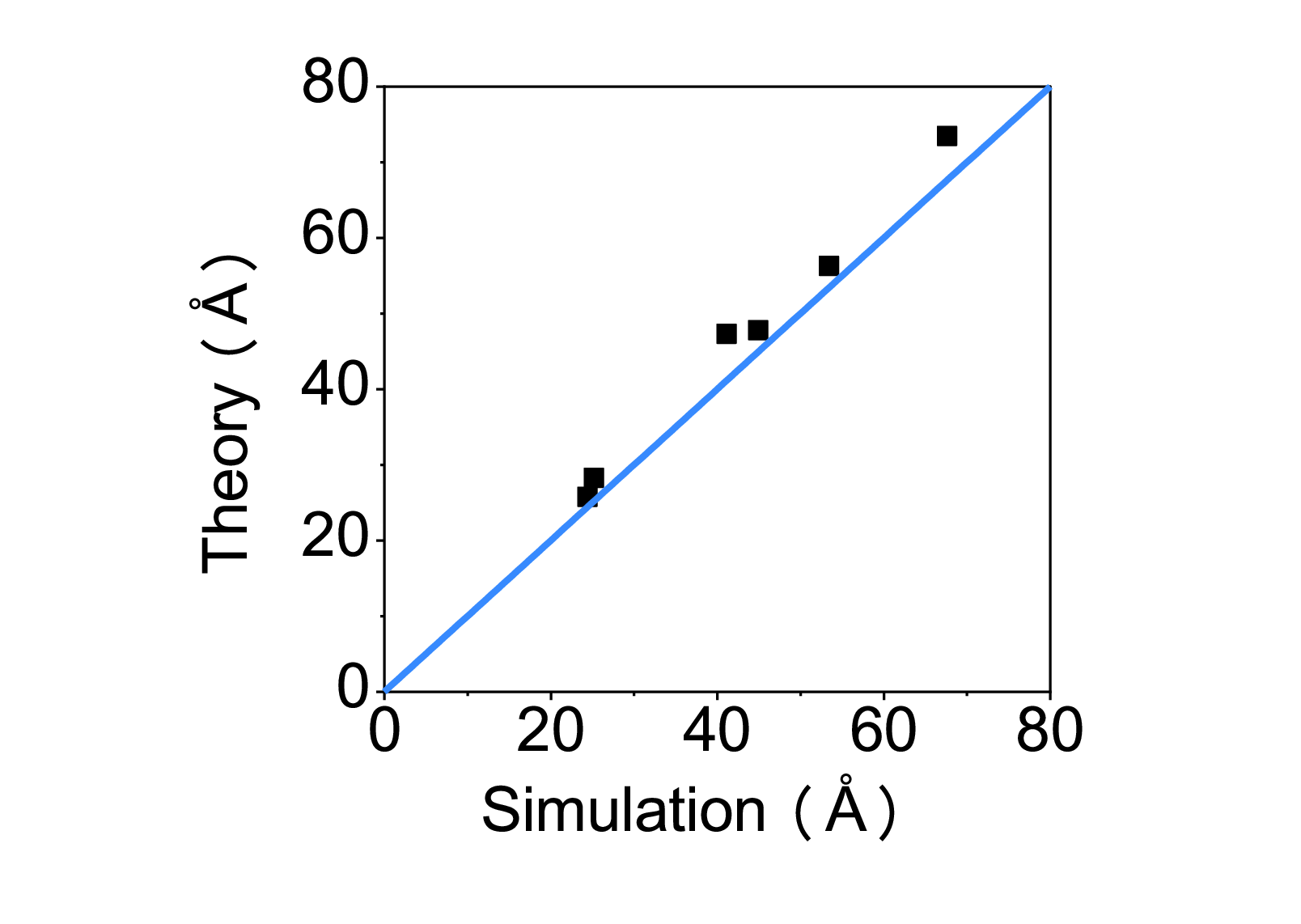}}
	\end{center}
	\caption{(Color online) Characteristic radius from numerical simulation versus analytic prediction by Eq.~(\ref{eq-radius}) for TMD Janus structures. The blue line represents $y=x$.}
	\label{fig_validation}
\end{figure}

In the first step, the S atom layer is stretched while the Te atom layer is compressed, so that the full structure containing three atomic layers remains in the planar configuration with the relaxed lattice constant $a_0$. The relaxed lattice constant can be obtained by minimizing the following total strain energy of S and Te atom layers,
\begin{eqnarray}
	V_{S} & = & \frac{1}{2}E_1(a_0/a_{10}-1)^2 + \frac{1}{2}E_2(a_0/a_{20}-1)^2,
	\label{eq_vs}
\end{eqnarray}
where $E_1$ and $E_2$ is the Young's modulus for MoS$_2$ and MoTe$_2$. The quantity $a_{10}$ and $a_{20}$ is the original lattice constant for MoS$_2$ and MoTe$_2$, respectively. The relaxed lattice constant $a_0$ obtained from minimizing the strain energy in Eq.~(\ref{eq_vs}) is
\begin{eqnarray}
	a_0 & = & \frac{E_1 a_{10} + E_2 a_{20}}{E_1 + E_2}.
	\label{eq_a0}
\end{eqnarray}
If $E_1=E_2$, Eq.~(\ref{eq_a0}) will be reduced to
\begin{eqnarray}
	a_0 & = & \frac{a_{10} + a_{20}}{2}.
\end{eqnarray}

In the second step, the stretching strain energy in the S atom layer and the compressive strain energy in the Te atom layer is released by bending the MoSTe monolayer (with S atoms in the inner surface and Te atoms in the outer surface). Assuming the bending structure of the MoSTe monolayer is a cylindrical surface with radius $r$, the total strain energy is
\begin{eqnarray}
	V_{\rm total} & = & \frac{1}{2}E_{1}\left[\left(1+\frac{h}{r}\right)-a_{10}/a_{0}\right]^{2} + \frac{1}{2}E_{2}\left[\left(1-\frac{h}{r}\right)-a_{20}/a_{0}\right]^{2},
\end{eqnarray}
where it has been assumed that the middle Mo atom layer is the neutral plane during the bending process.

The characteristic radius $R_C$ for equilibrium configuration of the MoSTe monolayer is achieved by
\begin{equation}
	\frac{\partial V_{\rm total}}{\partial r}|_{r=R_C}  =  0,
\end{equation}
which gives 
\begin{eqnarray}
	\frac{1}{R_C} & = & \frac{1}{(E_{1}+E_{2})ha_{0}}[(E_{1}a_{10}-E_{2}a_{20})-(E_{1}-E_{2})a_{0}].
	\label{eq-radius1}
\end{eqnarray}
If $E_1=E_2$, Eq.~(\ref{eq-radius1}) will be simplified to
\begin{equation}
	R_C = h\frac{a_{10}+a_{20}}{a_{10}-a_{20}}.
	\label{eq-radius}
\end{equation}

As listed in Tab.~(\ref{tab_Ea}), the difference between the Young's modulus of TMD materials is not large. The simplified analytic formula in Eq.~(\ref{eq-radius}) is applicable to all TMD Janus structures, though we have used MoSTe Janus monolayer as an explicit example during the derivation. This formula is used to predict the characteristic radius for these six TMD Janus structures MX$_2$ with (M = Mo, W, and X = S, Se, Te). Fig.~\ref{fig_validation} shows that the analytic prediction agrees quite well with the numerical simulation results.

\begin{table}[htbp]
	\centering
	\caption{The Young's modulus and lattice constant of six TMD monolayers.\cite{jiang2017handbook}}
	\label{tab_Ea}
	\begin{tabular}{c @{\hspace{2em}} c @{\hspace{2em}} c}
		\hline
		Structure &   Young's modulus (N/m)   & lattice constant (\AA) \\
		\hline
		MoS$_2$ & 97 & 4.13 \\
		MoSe$_2$ & 103.0 & 4.39 \\
		MoTe$_2$ & 79.8 & 4.73 \\
		WS$_2$ & 121.5 & 4.14 \\
		WSe$_2$ & 124.1 & 4.35 \\
		WTe$_2$ & 82.7 & 4.73 \\
		\hline
	\end{tabular}
\end{table}

\section{Abnormal phonon mode}

\begin{figure}[tb]
	\begin{center}
		\scalebox{1}[1]{\includegraphics[width=8cm]{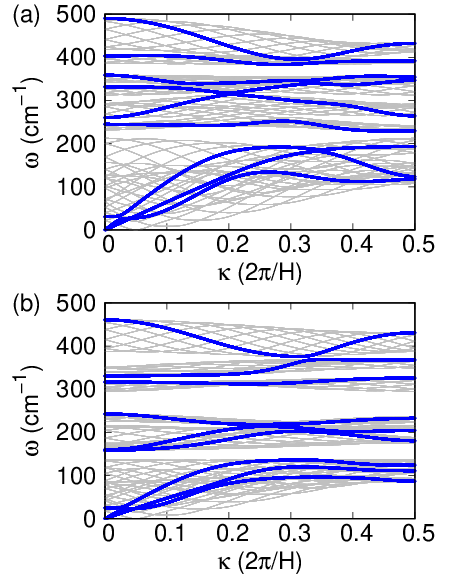}}
	\end{center}
	\caption{(Color online) Phonon dispersion for (a) MoS$_2$ and (b) Janus MoSTe (10, 10) nanotubes. Blue lines are for the special dispersions with the rotational wave vector $n=0$.}
	\label{fig_dispersion}
\end{figure}

\begin{figure*}[tb]
	\begin{center}
		\scalebox{1}[1]{\includegraphics[width=\textwidth]{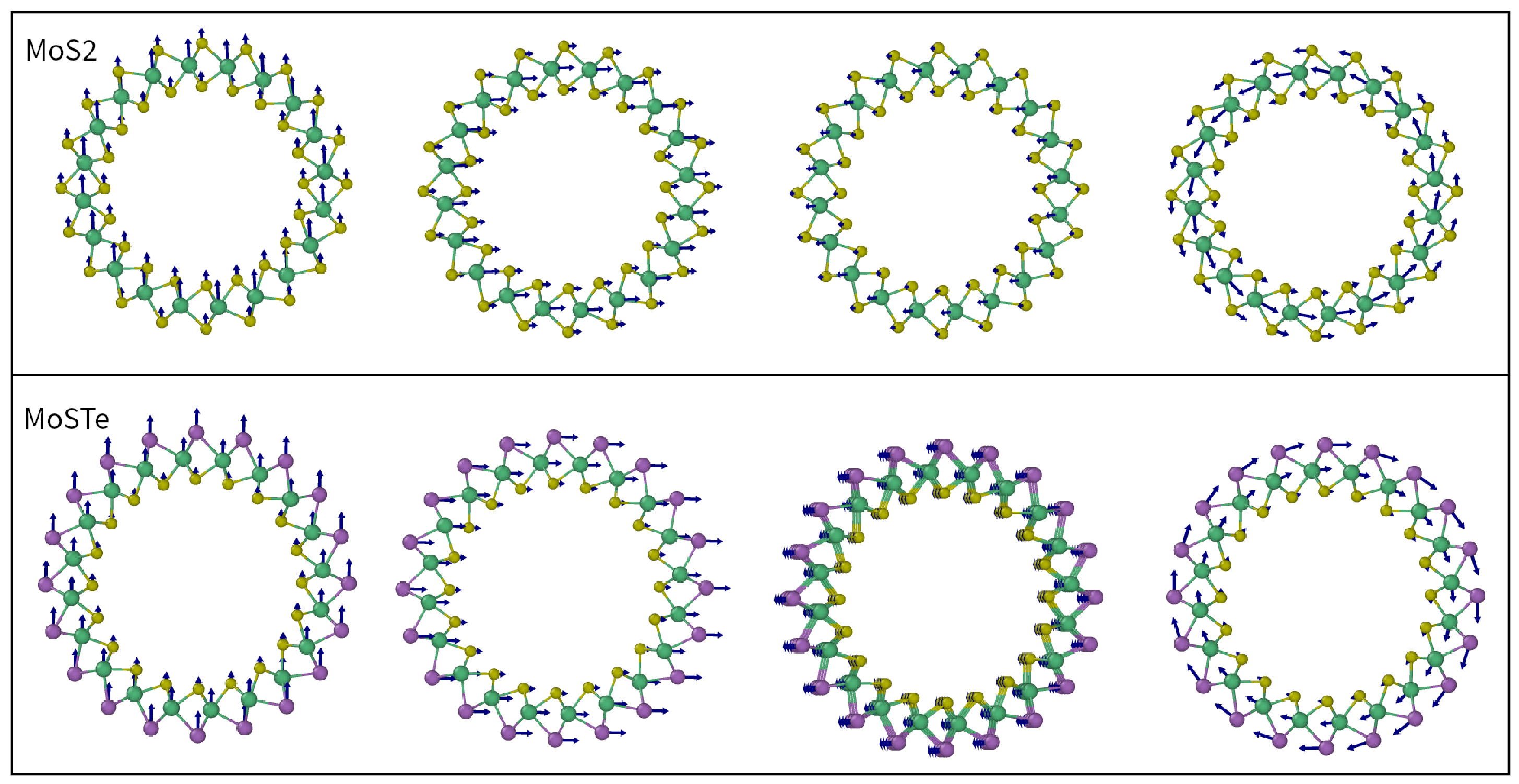}}
	\end{center}
	\caption{(Color online) Vibrational morphology for phonon modes with zero frequency in MoS$_2$ and Janus MoSTe (10, 10) nanotubes. Arrows correspond to the vibrational displacement of each atom in the mode.}
	\label{fig_u-zero}
\end{figure*}

\begin{figure}[tb]
	\begin{center}
		\scalebox{1}[1]{\includegraphics[width=8cm]{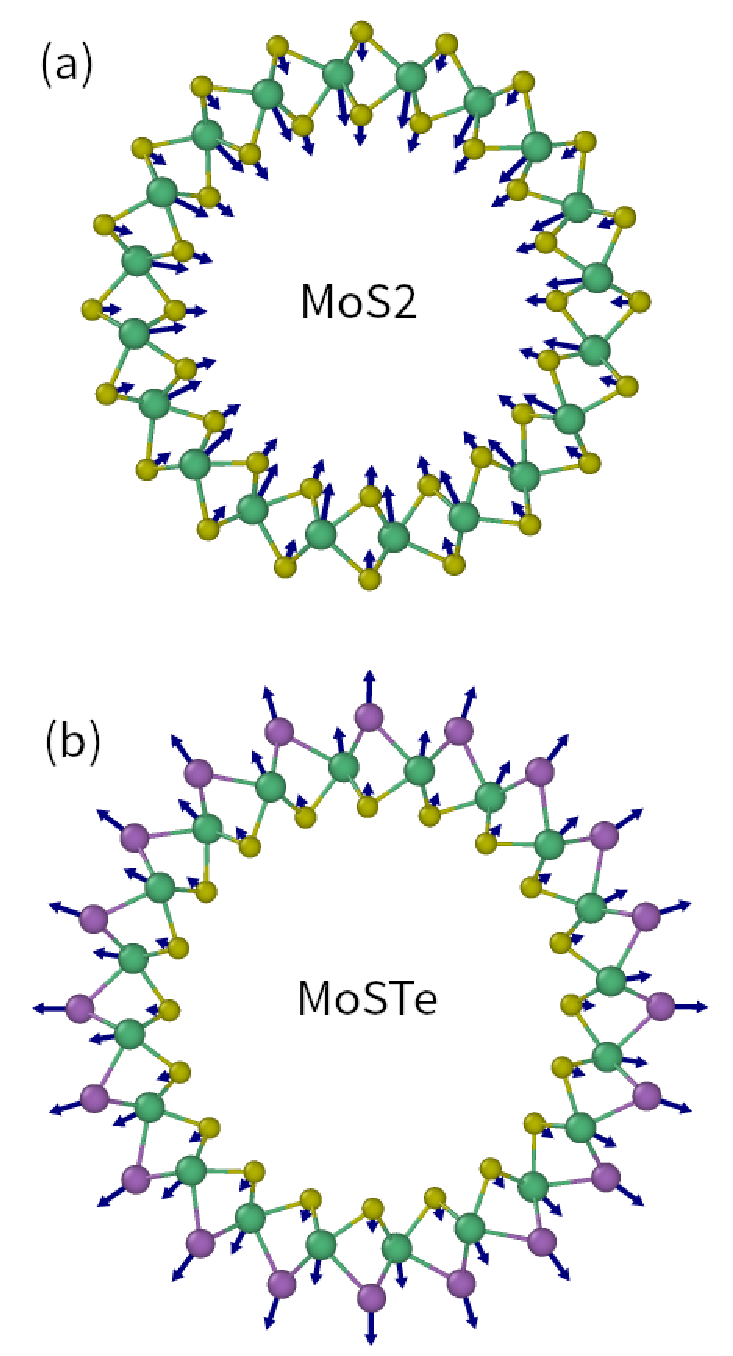}}
	\end{center}
	\caption{(Color online) Vibrational morphology for the radial breathing mode in (a) MoS$_2$ and (b) Janus MoSTe (10, 10) nanotubes.}
	\label{fig_u-rbm}
\end{figure}

\begin{figure}[tb]
	\begin{center}
		\scalebox{1}[1]{\includegraphics[width=8cm]{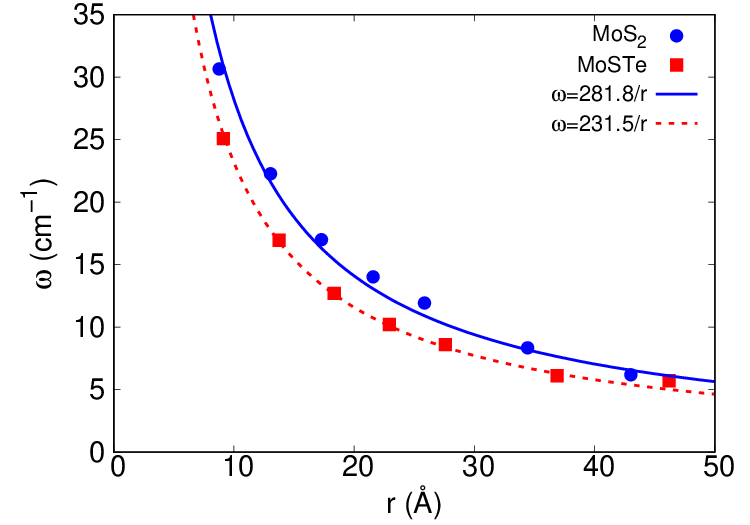}}
	\end{center}
	\caption{(Color online) Radial dependence for the radial breathing mode in MoS$_2$ and Janus MoSTe nanotubes.}
	\label{fig_rbm}
\end{figure}

In the previous section, we have explored that the TMD Janus nanotube with the characteristic radius is the most stable among all nanotubes with different radius. The phonon vibration mode is sensitive to the atomic configuration, so the characteristic radius may cause strong effects on the phonon modes of the TMD Janus nanotube. We will thus investigate the phonon modes for the TMD Janus nanotube with varying radius in this section.

The phonon dispersion for the MoS$_2$ nanotube and MoSTe nanotube are shown in Fig.~\ref{fig_dispersion}. The helical quantum numbers $(\kappa,n)$ are used, which correspond to the helical screw symmetry operation $\vec{R}_{H}$ and the rotational symmetry operation $\vec{R}_{n}$ in the nanotube.\cite{popov1999lattice,jiang2014phonon} The wave vector described by the helical quantum number is
\begin{eqnarray}
	\vec{k} & = & \kappa\vec{b}_{H}+n\vec{b}_{n},
\end{eqnarray}
where $\vec{b}_{H}$ and $\vec{b}_{n}$ are basic wave vectors in the reciprocal space, determined by the screw symmetry operation and rotational symmetry operation in the nanotube structure
\begin{eqnarray}
	\vec{b}_{H}\cdot\vec{R}_{H} & = & 2\pi,\\
	\vec{b}_{n}\cdot\vec{R}_{n} & = & 2\pi.
\end{eqnarray}
Screw and rotational operations construct the full symmetry group of the nanotube structure, which is named the line group.\cite{damnjanovic1999full,damnjanovic2000modified,tang2011symmetry} More details on the helical quantum numbers can be found in previous works on the lattice dynamics of single-walled nanotubes.\cite{popov1999lattice,popov2003lattice,jiang2014phonon}

\subsection{Acoustic phonon mode}

The phonon spectrum with quantum number $n=0$ have higher symmetry. In particular, we focus on these high symmetry phonon modes with helical quantum numbers $(\kappa,n)=(0,0)$, i.e., the phonon modes at $\Gamma$ point. There are four phonon modes with zero frequency in the nanotube structure, as shown in Fig.~\ref{fig_u-zero}, for both MoS$_2$ and MoSTe nanotubes. The longitudinal acoustic mode and the twisting mode are at the $\Gamma$ point. These two transverse acoustic modes are the flexural modes of the nanotube structure, which locate at a nonzero wave vector $\kappa$.\cite{popov1999lattice,JiangJW2006}

\subsection{Radial breathing mode}

The vibration morphology of the radial breathing mode (RBM) is shown in Fig.~\ref{fig_u-rbm}. The whole nanotube vibrates in a breathing-like mode in the RBM. The radial dependence for the frequency of the RBM is shown in Fig.~\ref{fig_rbm}. It shows that the frequency is inversely proportional to the radius of the nanotube, which has also be found in several previous works for similar hollow nanotube structures.\cite{kurti1998first,popov2006radius} The frequency of the MoSTe nanotube is obviously lower than the MoS$_2$ nanotube, although the elastic modulus has similar value in these two materials.\cite{jiang2017handbook} We will now explain the origin for the lower frequency of the RBM in the MoSTe nanotube as compared with the MoS$_2$ nanotube.

There are two energy terms that are involved in the vibration morphology of the RBM, including the stretching energy along the circumferential direction and the bending energy of the atomic layer. For the RBM, the displacement is along the radial direction
\begin{eqnarray}
	\vec{u} & = & \Delta r\hat{e}_{r}.
\end{eqnarray}
The vibration induced variation for the perimeter is
\begin{eqnarray}
	\Delta c & = & 2\pi\Delta r.
\end{eqnarray}
The resultant strain along the circumferential direction is
\begin{eqnarray}
	\epsilon & = & \frac{\Delta c}{c}=\frac{\Delta r}{r}.
\end{eqnarray}

The vibration induced variation in the stretching energy is
\begin{eqnarray}
	\Delta V_{S} & = & \frac{1}{2}E\epsilon^{2}=\frac{1}{2}\frac{E}{r^{2}}\left(\Delta r\right)^{2}.
\end{eqnarray}
The effective force constant from the stretching energy is 
\begin{eqnarray}
	K_{S} & = & \frac{E}{r^{2}}.
\end{eqnarray}

The bending energy is different for the MoS$_2$ nanotube and the MoSTe nanotube. For MoS$_2$, according to Eq.~(\ref{eq_vb_pure}), the variation of the bending energy due to the vibration of the RBM is
\begin{eqnarray}
	\Delta V_{B} & = & \frac{1}{2}\frac{\partial^{2}V_{B}}{\partial r^{2}}\left(\Delta r\right)^{2}=\frac{1}{2}\left(3D\frac{1}{r^{4}}\right)\left(\Delta r\right)^{2}.
\end{eqnarray}
The effective force constant of the bending energy is
\begin{eqnarray}
	K_{B} & = & 3D\frac{1}{r^{4}}.
\end{eqnarray}
The effective force constant for the RBM in the MoS$_2$ is
\begin{eqnarray}
	K_{\rm RBM} & = & K_{S}+K_{B}=\frac{E}{r^{2}}+3D\frac{1}{r^{4}}.
\end{eqnarray}
As a result, the radial dependence for the frequency of RBM is\cite{BornM}
\begin{eqnarray}
	\omega & = & 2\pi\sqrt{\frac{K_{\rm RBM}}{\rho h}}\approx2\pi\sqrt{\frac{E}{\rho h}}\frac{1}{r}\left(1+\frac{3}{2}\frac{D}{E}\frac{1}{r^{2}}\right).
\end{eqnarray}
The first term is $\frac{1}{r}$. This is exactly what we have found in the numerical simulation.

For MoSTe, the variation of the bending energy according to Eq.~(\ref{eq_vb_janus}) is
\begin{eqnarray}
	\Delta V_{B} & = & \frac{1}{2}\frac{\partial^{2}V_{B}}{\partial r^{2}}\left(\Delta r\right)^{2}=\frac{1}{2}D\left(\frac{3}{r^{4}}-\frac{2}{R_{C}r^{3}}\right)\left(\Delta r\right)^{2}.
\end{eqnarray}
As a result, the effective force constant for the RBM in MoSTe is
\begin{eqnarray}
	K_{\rm RBM} & = & K_{S}+K_{B}=\frac{E}{r^{2}}+3D\frac{1}{r^{4}}-\frac{2D}{R_{C}r^{3}}.\end{eqnarray}

The radial dependence for the frequency of RBM is
\begin{eqnarray}
	\omega & = & 2\pi\sqrt{\frac{K_{\rm RBM}}{\rho h}}\approx2\pi\sqrt{\frac{E}{\rho h}}\frac{1}{r}\left(1+\frac{3}{2}\frac{D}{E}\frac{1}{r^{2}}-\frac{2D}{ER_{C}}\frac{1}{r}\right).
\end{eqnarray}
We find that the correction in the bending energy contributes a negative component with $r^{-2}$ radial dependence. This is one of the origins for the lower frequency of the MoSTe nanotube as compared with the MoS$_2$ nanotube.

\subsection{Optical phonon mode in MoS$_2$ nanotube}

\begin{figure*}[tb]
	\begin{center}
		\scalebox{1}[1]{\includegraphics[width=\textwidth]{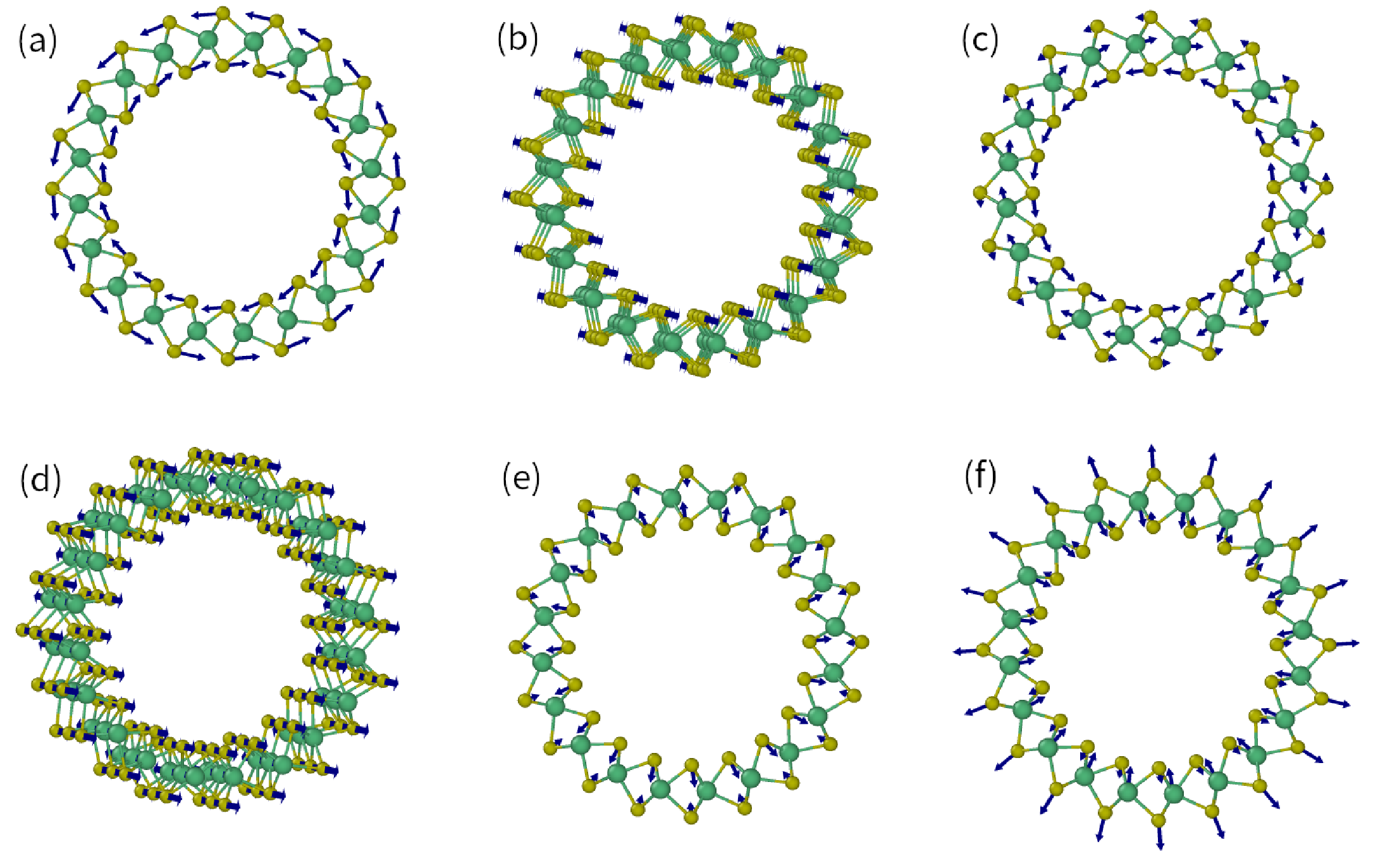}}
	\end{center}
	\caption{(Color online) Vibrational morphology for these six optical phonon modes at the $\Gamma$ point with wave vector $(\kappa, n)=(0,0)$ in (10, 10) MoS$_2$ nanotube.}
	\label{fig_u-optical-mos2}
\end{figure*}

\begin{figure*}[tb]
	\begin{center}
		\scalebox{1}[1]{\includegraphics[width=\textwidth]{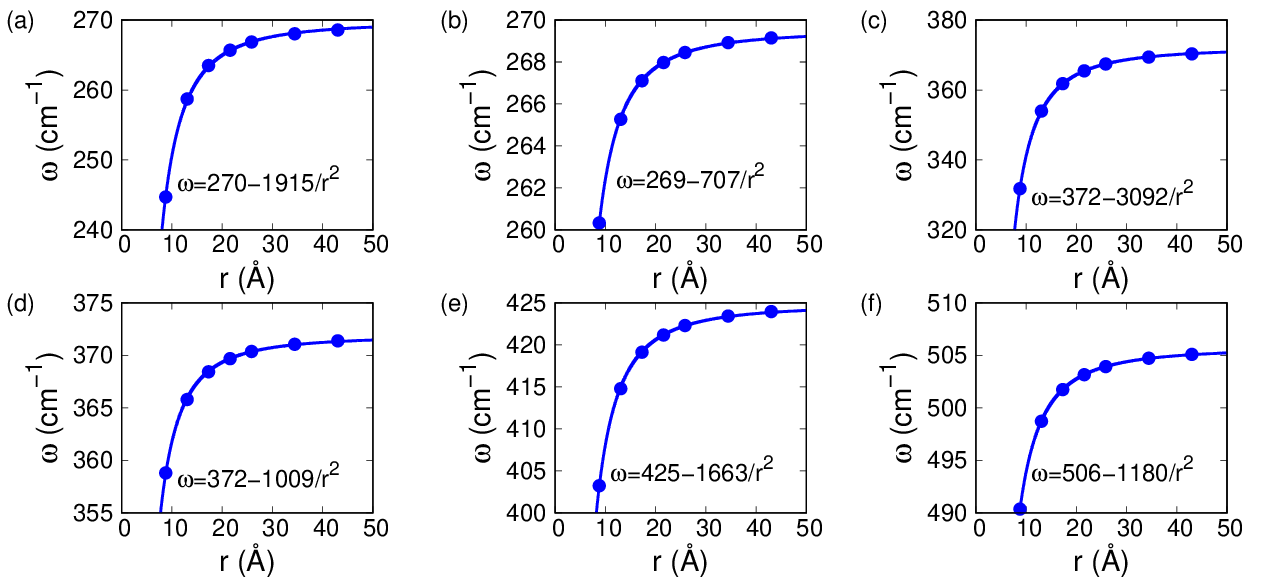}}
	\end{center}
	\caption{(Color online) Radial dependence for these six optical phonon modes in MoS$_2$ nanotubes as shown in Fig.~\ref{fig_u-optical-mos2}.}
	\label{fig_phonon-mos2}
\end{figure*}

\begin{figure*}[tb]
	\begin{center}
		\scalebox{1}[1]{\includegraphics[width=\textwidth]{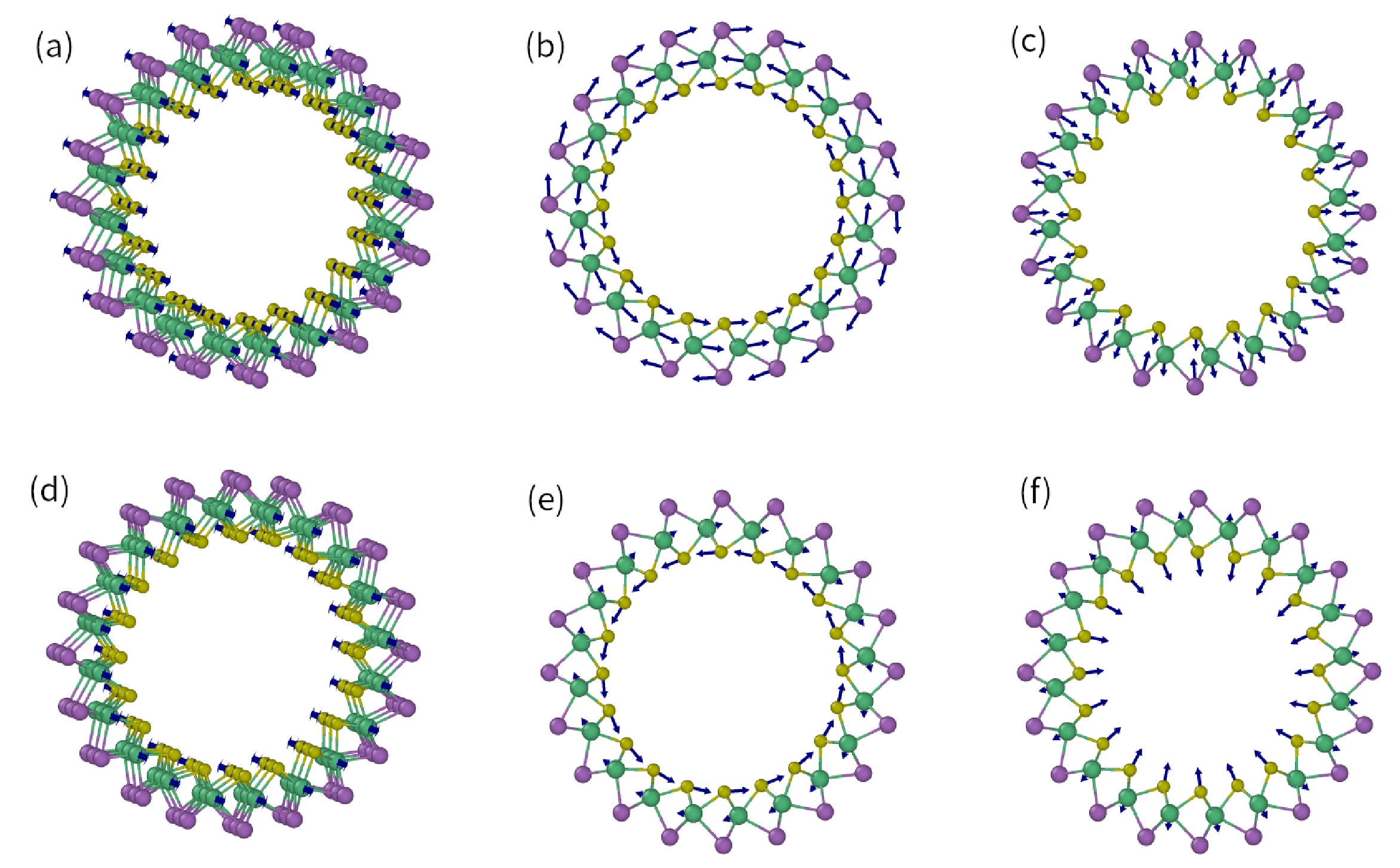}}
	\end{center}
	\caption{(Color online) Vibrational morphology for these six optical phonon modes at the $\Gamma$ point with wave vector $(\kappa, n)=(0,0)$ in (10, 10) the Janus MoSTe nanotube.}
	\label{fig_u-optical-moste}
\end{figure*}

\begin{figure*}[tb]
	\begin{center}
		\scalebox{1}[1]{\includegraphics[width=\textwidth]{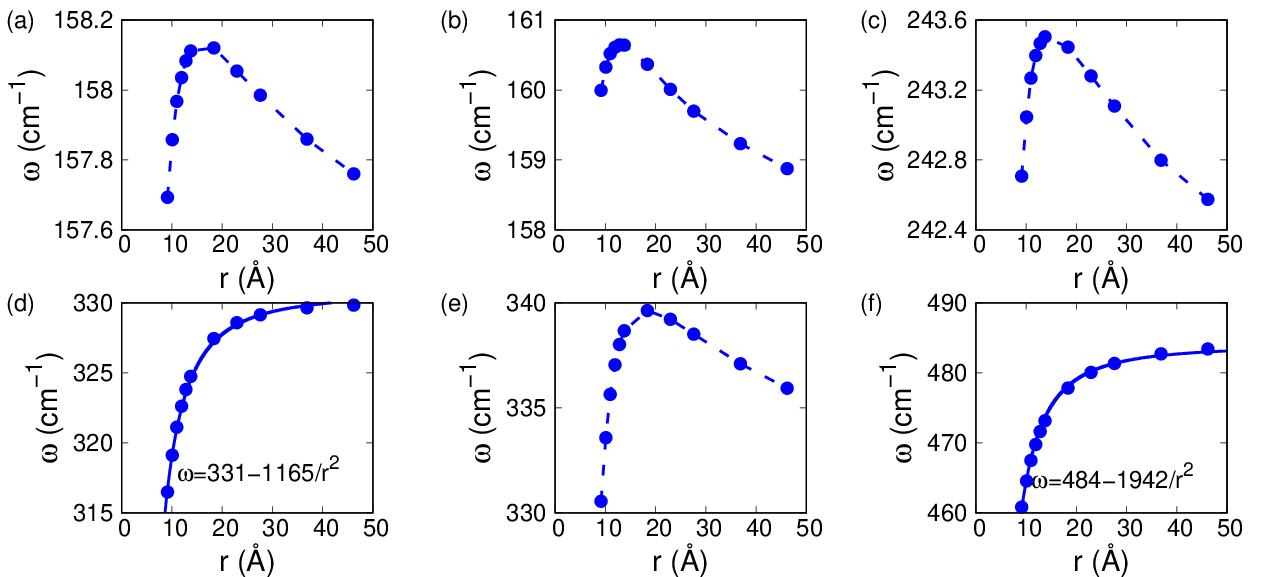}}
	\end{center}
	\caption{(Color online) Radial dependence for these six optical phonon modes in Janus MoSTe nanotubes as shown in Fig.~\ref{fig_u-optical-moste}.}
	\label{fig_phonon-moste}
\end{figure*}

We now study these six optical phonon modes at the $\Gamma$ point with $(\kappa,n)=(0,0)$. The vibration morphology of these six optical phonon modes in the MoS$_2$ nanotube are shown in Fig.~\ref{fig_u-optical-mos2}. The corresponding radial dependence of the frequency for each optical phonon mode is shown in Fig.~\ref{fig_phonon-mos2}. It shows that the frequency of the optical phonon mode increases with increasing tube radius as a function $r^{-2}$, instead of $r^{-1}$ for RBM.

The radial dependence is closely related to the vibration morphology of the optical phonon mode. For these six optical phonon modes, there are four shear optical modes within the MoS$_2$ atomic layer, as shown in Fig.~\ref{fig_u-optical-mos2}~(a)-(d). There are two stretching optical modes along the perpendicular direction of the atomic layer, as shown in Fig.~\ref{fig_u-optical-mos2}~(e) and (f). The vibration of these four optical shear modes are governed by the shear energy\cite{LandauLD}
\begin{eqnarray}
	V_{G} & = & \frac{1}{2}G\gamma^{2},
\end{eqnarray}
with $\gamma$ as the shear strain and $G$ as the shear modulus of the atomic layer.

For nanotubes, the shear strain needs to be projected onto the tangential (circumferential or axial) direction
\begin{eqnarray}
	\gamma' & = & \gamma\cos\theta\approx\gamma\left[1-\frac{1}{2}\left(\frac{b}{r}\right)^{2}\right],
\end{eqnarray}
with $\theta$ as the tubular curvature induced tangential angle, and $b$ is the bond length.

The shear energy density in the MoS$_2$ nanotube is
\begin{eqnarray}
	V_{G} & = & \frac{1}{2}G\gamma'{}^{2}=\frac{1}{2}G\left[1-\frac{1}{2}\left(\frac{b}{r}\right)^{2}\right]^{2}\gamma^{2},
\end{eqnarray}
so the effective shear modulus in the nanotube is
\begin{eqnarray}
	G' & = & G\left[1-\frac{1}{2}\left(\frac{b}{r}\right)^{2}\right]^{2}.
\end{eqnarray}
The frequency for the shearing optical phonon mode is
\begin{eqnarray}
	\omega & = & 2\pi\sqrt{\frac{G'}{\rho h}}=2\pi\sqrt{\frac{G}{\rho h}}\times\left[1-\frac{1}{2}\left(\frac{b}{r}\right)^{2}\right]=\omega_{0}\times\left[1-\frac{1}{2}\left(\frac{b}{r}\right)^{2}\right].
\end{eqnarray}
It shows that the frequency depends on the radius as $r^{-2}$, instead of $r^{-1}$ in the RBM. Indeed, the fitting error will be obviously increased if we use $r^{-1}$ to fit the numerical simulation data.

\subsection{Optical phonon mode in MoSTe nanotube}

For MoSTe, the vibration morphology for these six optical modes are shown in Fig.~\ref{fig_u-optical-moste}. The corresponding radial dependence for the frequency of these optical modes are shown in Fig.~\ref{fig_phonon-moste}. There are two major different features for the radial dependence for the optical modes in MoSTe, as compared with MoS$_2$. Firstly, the frequency for the first 3 optical modes only vary slightly with varying radius. Secondly, the radial dependence for these four optical modes in Fig.~\ref{fig_phonon-moste}~(a), (b), (c), and (e) modes have a maximum point, instead of monotonic increase with increasing radius in the MoS$_2$ nanotube.

The first feature can be explained as follows. In these first three optical modes, all atoms (S, Mo, Te) are involved in the vibration morphology. The Te atom is the heaviest atom. The S and Mo atoms are neighboring atoms. The vibration direction of S and Mo atoms are in the same direction, so the relative vibration between S and Mo atom is neglectable and has no radial dependence. As a result, the frequency of the first three modes in the MoSTe nanotube are not sensitive to the radius. The situation in MoS$_2$ is different, where all neighboring atoms have opposite vibration directions in the optical phonon modes.

The second feature is related to the characteristic radius of the MoSTe Janus nanotube. For the MoSTe nanotube with the critical characteristic radius $R_{C}$, the structure is at the energy minimum configuration. For MoSTe nanotube with other radius different from $R_{C}$, the structure deviates from the most stable configuration, which will result in the soft mode effect. The soft mode is the phonon mode that drives the structure to transit toward the most stable configuration. The effect of the vibration for the soft mode is to reduce the frequency, leading to low frequency and even imaginary frequency of the soft mode.\cite{scott1974soft,nakanishi1982lattice,rudin2018generalization,gupta2022soft,pallikara2022physical} For the soft mode with eigen vector $e_{i\alpha}$, the atomic displacement for atom i along the $\alpha$ direction can be expanded as\cite{BornM}
\begin{eqnarray}
	u_{i\alpha} & = & \frac{e_{i\alpha}}{\sqrt{M_{i}}}Q,
\end{eqnarray}
with $Q$ as the coordinate in the mode space.

The total vibration energy can be expanded in terms of the vibration morphology for the soft mode\cite{parlinski1997first,baroni2001phonons,togo2008first}
\begin{eqnarray}
	V\left(Q\right) & = & V_{0}+\frac{1}{2}K_{Q}^{(2)}Q^{2}+O\left(Q^{4}\right),
\end{eqnarray}
with $K_{Q}^{(2)}$ as the force constant parameter. This parameter is a negative value for the soft mode, leading to the imaginary frequency since $\omega \propto \sqrt{K_{Q}^{(2)}}$.

The potential energy difference between the MoSTe nanotube with radius $r$ and the most stable MoSTe nanotube with radius $R_C$ is mainly contributed by the bending energy
\begin{eqnarray}
	\Delta V_{C} & = & V_{B}(r)-V_{B}(R_{C})=\frac{1}{2}D\frac{\left(r-R_{C}\right)^{2}}{r^{2}R_{C}^{2}}.
\end{eqnarray}
The soft mode effect is proportional to the magnitude of the structure deviation from the most stable configuration. As a result, the soft mode induced reduction of the frequency is\cite{cochran1960crystal,cowley1964lattice,zhong1995first}
\begin{eqnarray}
	\Delta\omega & \propto & -\sqrt{\Delta V_{C}}=-\sqrt{\frac{1}{2}D}|\frac{1}{r}-\frac{1}{R_{C}}|.
\end{eqnarray}
It shows that, due to the soft mode effect, the frequency will be reduced with increasing radius for $r>R_{C}$, which will compete with the curvature effect discussed above. As a result, there is a maximum frequency around the critical characteristic radius $R_{C}$.

It should be noted that the soft mode effect has the strongest effect on the RBM, because the vibration morphology of the RBM tries to vary the radius of the MoSTe nanotube, driving the structure toward the most stable configuration. Hence, the soft mode effect is another important origin that reduces the frequency of the RBM in the MoSTe nanotube, as compared with the MoS$_2$ nanotube.

For the other optical modes, the soft mode can take effect through mode coupling or through the relevance of the eigen vectors for the optical mode and the soft mode. This soft mode effect thus serves as the origin for the abnormal radial dependence of these four optical modes in Fig.~\ref{fig_u-optical-moste}~(a), (b), (c), and (e). However, for these two optical modes in Fig.~\ref{fig_u-optical-moste}~(d) and (f), the soft mode effect is weaker than the previous curvature induced projection effect, so the frequency increases with increasing radius.

\section{conclusion}

In summary, we have systematically studied the structural stability and vibrational properties of TMD Janus nanotubes obtained by rolling Janus monolayers. We revealed the existence of a characteristic radius at which the nanotube achieves its lowest total energy, establishing the most stable geometry. A continuum mechanics model was developed to derive the analytical expression for this radius, offering physical insight into the characteristic radius. Furthermore, our analysis demonstrated that the optical phonon modes of Janus nanotubes exhibit an anomalous non-monotonic dependence on tube radius, with a distinct maximum frequency emerging near the characteristic radius. This behavior distinguishes Janus nanotubes from conventional tubular systems and highlights the critical role of curvature coupling in shaping their fundamental properties. Our findings not only advance the theoretical understanding of Janus-based curved nanostructures but also suggest new opportunities for engineering phononic and electronic functionalities in next-generation nanomaterials.

\textbf{Acknowledgment} The first author thanks Yu Li at Shanghai University for numerical simulation supports. This work is supported by the National Natural Science Foundation of China (Grant Nos. 12072182 and 12421002).

\textbf{Data Availability} The data supporting this study's findings are available within the article.



\begin{thebibliography}{61}%
	\makeatletter
	\providecommand \@ifxundefined [1]{%
		\@ifx{#1\undefined}
	}%
	\providecommand \@ifnum [1]{%
		\ifnum #1\expandafter \@firstoftwo
		\else \expandafter \@secondoftwo
		\fi
	}%
	\providecommand \@ifx [1]{%
		\ifx #1\expandafter \@firstoftwo
		\else \expandafter \@secondoftwo
		\fi
	}%
	\providecommand \natexlab [1]{#1}%
	\providecommand \enquote  [1]{``#1''}%
	\providecommand \bibnamefont  [1]{#1}%
	\providecommand \bibfnamefont [1]{#1}%
	\providecommand \citenamefont [1]{#1}%
	\providecommand \href@noop [0]{\@secondoftwo}%
	\providecommand \href [0]{\begingroup \@sanitize@url \@href}%
	\providecommand \@href[1]{\@@startlink{#1}\@@href}%
	\providecommand \@@href[1]{\endgroup#1\@@endlink}%
	\providecommand \@sanitize@url [0]{\catcode `\\12\catcode `\$12\catcode
		`\&12\catcode `\#12\catcode `\^12\catcode `\_12\catcode `\%12\relax}%
	\providecommand \@@startlink[1]{}%
	\providecommand \@@endlink[0]{}%
	\providecommand \url  [0]{\begingroup\@sanitize@url \@url }%
	\providecommand \@url [1]{\endgroup\@href {#1}{\urlprefix }}%
	\providecommand \urlprefix  [0]{URL }%
	\providecommand \Eprint [0]{\href }%
	\providecommand \doibase [0]{http://dx.doi.org/}%
	\providecommand \selectlanguage [0]{\@gobble}%
	\providecommand \bibinfo  [0]{\@secondoftwo}%
	\providecommand \bibfield  [0]{\@secondoftwo}%
	\providecommand \translation [1]{[#1]}%
	\providecommand \BibitemOpen [0]{}%
	\providecommand \bibitemStop [0]{}%
	\providecommand \bibitemNoStop [0]{.\EOS\space}%
	\providecommand \EOS [0]{\spacefactor3000\relax}%
	\providecommand \BibitemShut  [1]{\csname bibitem#1\endcsname}%
	\let\auto@bib@innerbib\@empty
	\bibitem [{\citenamefont {Manzeli}\ \emph {et~al.}(2017)\citenamefont
		{Manzeli}, \citenamefont {Ovchinnikov}, \citenamefont {Pasquier},
		\citenamefont {Yazyev},\ and\ \citenamefont {Kis}}]{manzeli20172d}%
	\BibitemOpen
	\bibfield  {author} {\bibinfo {author} {\bibfnamefont {S.}~\bibnamefont
			{Manzeli}}, \bibinfo {author} {\bibfnamefont {D.}~\bibnamefont
			{Ovchinnikov}}, \bibinfo {author} {\bibfnamefont {D.}~\bibnamefont
			{Pasquier}}, \bibinfo {author} {\bibfnamefont {O.~V.}\ \bibnamefont
			{Yazyev}}, \ and\ \bibinfo {author} {\bibfnamefont {A.}~\bibnamefont {Kis}},\
	}\href@noop {} {\bibfield  {journal} {\bibinfo  {journal} {Nature Reviews
				Materials}\ }\textbf {\bibinfo {volume} {2}},\ \bibinfo {pages} {17033}
		(\bibinfo {year} {2017})}\BibitemShut {NoStop}%
	\bibitem [{\citenamefont {Choi}\ \emph {et~al.}(2017)\citenamefont {Choi},
		\citenamefont {Choudhary}, \citenamefont {Han}, \citenamefont {Park},
		\citenamefont {Akinwande},\ and\ \citenamefont {Lee}}]{choi2017recent}%
	\BibitemOpen
	\bibfield  {author} {\bibinfo {author} {\bibfnamefont {W.}~\bibnamefont
			{Choi}}, \bibinfo {author} {\bibfnamefont {N.}~\bibnamefont {Choudhary}},
		\bibinfo {author} {\bibfnamefont {G.~H.}\ \bibnamefont {Han}}, \bibinfo
		{author} {\bibfnamefont {J.}~\bibnamefont {Park}}, \bibinfo {author}
		{\bibfnamefont {D.}~\bibnamefont {Akinwande}}, \ and\ \bibinfo {author}
		{\bibfnamefont {Y.~H.}\ \bibnamefont {Lee}},\ }\href@noop {} {\bibfield
		{journal} {\bibinfo  {journal} {Materials Today}\ }\textbf {\bibinfo {volume}
			{20}},\ \bibinfo {pages} {116} (\bibinfo {year} {2017})}\BibitemShut
	{NoStop}%
	\bibitem [{\citenamefont {Wang}\ \emph {et~al.}(2024)\citenamefont {Wang},
		\citenamefont {Sarkar}, \citenamefont {Yan},\ and\ \citenamefont
		{Chhowalla}}]{wang2024critical}%
	\BibitemOpen
	\bibfield  {author} {\bibinfo {author} {\bibfnamefont {Y.}~\bibnamefont
			{Wang}}, \bibinfo {author} {\bibfnamefont {S.}~\bibnamefont {Sarkar}},
		\bibinfo {author} {\bibfnamefont {H.}~\bibnamefont {Yan}}, \ and\ \bibinfo
		{author} {\bibfnamefont {M.}~\bibnamefont {Chhowalla}},\ }\href@noop {}
	{\bibfield  {journal} {\bibinfo  {journal} {Nature Electronics}\ }\textbf
		{\bibinfo {volume} {7}},\ \bibinfo {pages} {638} (\bibinfo {year}
		{2024})}\BibitemShut {NoStop}%
	\bibitem [{\citenamefont {Gupta}\ \emph {et~al.}(2025)\citenamefont {Gupta},
		\citenamefont {Zhang}, \citenamefont {Lei}, \citenamefont {Yu}, \citenamefont
		{Liu}, \citenamefont {Zou},\ and\ \citenamefont {Yakobson}}]{gupta2025two}%
	\BibitemOpen
	\bibfield  {author} {\bibinfo {author} {\bibfnamefont {S.}~\bibnamefont
			{Gupta}}, \bibinfo {author} {\bibfnamefont {J.-J.}\ \bibnamefont {Zhang}},
		\bibinfo {author} {\bibfnamefont {J.}~\bibnamefont {Lei}}, \bibinfo {author}
		{\bibfnamefont {H.}~\bibnamefont {Yu}}, \bibinfo {author} {\bibfnamefont
			{M.}~\bibnamefont {Liu}}, \bibinfo {author} {\bibfnamefont {X.}~\bibnamefont
			{Zou}}, \ and\ \bibinfo {author} {\bibfnamefont {B.~I.}\ \bibnamefont
			{Yakobson}},\ }\href@noop {} {\bibfield  {journal} {\bibinfo  {journal}
			{Chemical Reviews}\ }\textbf {\bibinfo {volume} {125}},\ \bibinfo {pages}
		{786} (\bibinfo {year} {2025})}\BibitemShut {NoStop}%
	\bibitem [{\citenamefont {Han}\ \emph {et~al.}(2018)\citenamefont {Han},
		\citenamefont {Duong}, \citenamefont {Keum}, \citenamefont {Yun},\ and\
		\citenamefont {Lee}}]{han2018van}%
	\BibitemOpen
	\bibfield  {author} {\bibinfo {author} {\bibfnamefont {G.~H.}\ \bibnamefont
			{Han}}, \bibinfo {author} {\bibfnamefont {D.~L.}\ \bibnamefont {Duong}},
		\bibinfo {author} {\bibfnamefont {D.~H.}\ \bibnamefont {Keum}}, \bibinfo
		{author} {\bibfnamefont {S.~J.}\ \bibnamefont {Yun}}, \ and\ \bibinfo
		{author} {\bibfnamefont {Y.~H.}\ \bibnamefont {Lee}},\ }\href@noop {}
	{\bibfield  {journal} {\bibinfo  {journal} {Chemical Reviews}\ }\textbf
		{\bibinfo {volume} {118}},\ \bibinfo {pages} {6297} (\bibinfo {year}
		{2018})}\BibitemShut {NoStop}%
	\bibitem [{\citenamefont {Ma}, \citenamefont {Wang},\ and\ \citenamefont
		{Liu}(2024)}]{ma2024van}%
	\BibitemOpen
	\bibfield  {author} {\bibinfo {author} {\bibfnamefont {L.}~\bibnamefont
			{Ma}}, \bibinfo {author} {\bibfnamefont {Y.}~\bibnamefont {Wang}}, \ and\
		\bibinfo {author} {\bibfnamefont {Y.}~\bibnamefont {Liu}},\ }\href@noop {}
	{\bibfield  {journal} {\bibinfo  {journal} {Chemical Reviews}\ }\textbf
		{\bibinfo {volume} {124}},\ \bibinfo {pages} {2583} (\bibinfo {year}
		{2024})}\BibitemShut {NoStop}%
	\bibitem [{\citenamefont {Duan}\ \emph {et~al.}(2014)\citenamefont {Duan},
		\citenamefont {Wang}, \citenamefont {Shaw}, \citenamefont {Cheng},
		\citenamefont {Chen}, \citenamefont {Li}, \citenamefont {Wu}, \citenamefont
		{Tang}, \citenamefont {Zhang}, \citenamefont {Pan}, \citenamefont {Jiang},
		\citenamefont {Yu}, \citenamefont {Huang},\ and\ \citenamefont
		{Duan}}]{DuanX2014nn}%
	\BibitemOpen
	\bibfield  {author} {\bibinfo {author} {\bibfnamefont {X.}~\bibnamefont
			{Duan}}, \bibinfo {author} {\bibfnamefont {C.}~\bibnamefont {Wang}}, \bibinfo
		{author} {\bibfnamefont {J.~C.}\ \bibnamefont {Shaw}}, \bibinfo {author}
		{\bibfnamefont {R.}~\bibnamefont {Cheng}}, \bibinfo {author} {\bibfnamefont
			{Y.}~\bibnamefont {Chen}}, \bibinfo {author} {\bibfnamefont {H.}~\bibnamefont
			{Li}}, \bibinfo {author} {\bibfnamefont {X.}~\bibnamefont {Wu}}, \bibinfo
		{author} {\bibfnamefont {Y.}~\bibnamefont {Tang}}, \bibinfo {author}
		{\bibfnamefont {Q.}~\bibnamefont {Zhang}}, \bibinfo {author} {\bibfnamefont
			{A.}~\bibnamefont {Pan}}, \bibinfo {author} {\bibfnamefont {J.}~\bibnamefont
			{Jiang}}, \bibinfo {author} {\bibfnamefont {R.}~\bibnamefont {Yu}}, \bibinfo
		{author} {\bibfnamefont {Y.}~\bibnamefont {Huang}}, \ and\ \bibinfo {author}
		{\bibfnamefont {X.}~\bibnamefont {Duan}},\ }\href@noop {} {\bibfield
		{journal} {\bibinfo  {journal} {Nature Nanotechnology}\ }\textbf {\bibinfo
			{volume} {9}},\ \bibinfo {pages} {1024} (\bibinfo {year} {2014})}\BibitemShut
	{NoStop}%
	\bibitem [{\citenamefont {Wang}\ \emph {et~al.}(2019)\citenamefont {Wang},
		\citenamefont {Li}, \citenamefont {Chen}, \citenamefont {Deng},\ and\
		\citenamefont {Niu}}]{wang2019recent}%
	\BibitemOpen
	\bibfield  {author} {\bibinfo {author} {\bibfnamefont {J.}~\bibnamefont
			{Wang}}, \bibinfo {author} {\bibfnamefont {Z.}~\bibnamefont {Li}}, \bibinfo
		{author} {\bibfnamefont {H.}~\bibnamefont {Chen}}, \bibinfo {author}
		{\bibfnamefont {G.}~\bibnamefont {Deng}}, \ and\ \bibinfo {author}
		{\bibfnamefont {X.}~\bibnamefont {Niu}},\ }\href@noop {} {\bibfield
		{journal} {\bibinfo  {journal} {Nano-Micro Letters}\ }\textbf {\bibinfo
			{volume} {11}},\ \bibinfo {pages} {48} (\bibinfo {year} {2019})}\BibitemShut
	{NoStop}%
	\bibitem [{\citenamefont {Liu}\ \emph {et~al.}(2025)\citenamefont {Liu},
		\citenamefont {Li}, \citenamefont {Guo}, \citenamefont {Zhang}, \citenamefont
		{Shen}, \citenamefont {Ye}, \citenamefont {Peng}, \citenamefont {Qi},
		\citenamefont {Wu}, \citenamefont {Li} \emph {et~al.}}]{liu20252d}%
	\BibitemOpen
	\bibfield  {author} {\bibinfo {author} {\bibfnamefont {M.}~\bibnamefont
			{Liu}}, \bibinfo {author} {\bibfnamefont {S.}~\bibnamefont {Li}}, \bibinfo
		{author} {\bibfnamefont {Y.}~\bibnamefont {Guo}}, \bibinfo {author}
		{\bibfnamefont {L.}~\bibnamefont {Zhang}}, \bibinfo {author} {\bibfnamefont
			{D.}~\bibnamefont {Shen}}, \bibinfo {author} {\bibfnamefont {Q.}~\bibnamefont
			{Ye}}, \bibinfo {author} {\bibfnamefont {Z.}~\bibnamefont {Peng}}, \bibinfo
		{author} {\bibfnamefont {W.}~\bibnamefont {Qi}}, \bibinfo {author}
		{\bibfnamefont {R.}~\bibnamefont {Wu}}, \bibinfo {author} {\bibfnamefont
			{J.}~\bibnamefont {Li}},  \emph {et~al.},\ }\href@noop {} {\bibfield
		{journal} {\bibinfo  {journal} {Advanced Functional Materials}\ ,\ \bibinfo
			{pages} {2500876}} (\bibinfo {year} {2025})}\BibitemShut {NoStop}%
	\bibitem [{\citenamefont {Cheng}\ \emph {et~al.}(2013)\citenamefont {Cheng},
		\citenamefont {Zhu}, \citenamefont {Tahir},\ and\ \citenamefont
		{Schwingenschl{\"o}gl}}]{cheng2013spin}%
	\BibitemOpen
	\bibfield  {author} {\bibinfo {author} {\bibfnamefont {Y.}~\bibnamefont
			{Cheng}}, \bibinfo {author} {\bibfnamefont {Z.}~\bibnamefont {Zhu}}, \bibinfo
		{author} {\bibfnamefont {M.}~\bibnamefont {Tahir}}, \ and\ \bibinfo {author}
		{\bibfnamefont {U.}~\bibnamefont {Schwingenschl{\"o}gl}},\ }\href@noop {}
	{\bibfield  {journal} {\bibinfo  {journal} {Europhysics Letters}\ }\textbf
		{\bibinfo {volume} {102}},\ \bibinfo {pages} {57001} (\bibinfo {year}
		{2013})}\BibitemShut {NoStop}%
	\bibitem [{\citenamefont {Lu}\ \emph {et~al.}(2017)\citenamefont {Lu},
		\citenamefont {Zhu}, \citenamefont {Xiao}, \citenamefont {Chuu},
		\citenamefont {Han}, \citenamefont {Chiu}, \citenamefont {Cheng},
		\citenamefont {Yang}, \citenamefont {Wei}, \citenamefont {Yang} \emph
		{et~al.}}]{lu2017janus}%
	\BibitemOpen
	\bibfield  {author} {\bibinfo {author} {\bibfnamefont {A.-Y.}\ \bibnamefont
			{Lu}}, \bibinfo {author} {\bibfnamefont {H.}~\bibnamefont {Zhu}}, \bibinfo
		{author} {\bibfnamefont {J.}~\bibnamefont {Xiao}}, \bibinfo {author}
		{\bibfnamefont {C.-P.}\ \bibnamefont {Chuu}}, \bibinfo {author}
		{\bibfnamefont {Y.}~\bibnamefont {Han}}, \bibinfo {author} {\bibfnamefont
			{M.-H.}\ \bibnamefont {Chiu}}, \bibinfo {author} {\bibfnamefont {C.-C.}\
			\bibnamefont {Cheng}}, \bibinfo {author} {\bibfnamefont {C.-W.}\ \bibnamefont
			{Yang}}, \bibinfo {author} {\bibfnamefont {K.-H.}\ \bibnamefont {Wei}},
		\bibinfo {author} {\bibfnamefont {Y.}~\bibnamefont {Yang}},  \emph {et~al.},\
	}\href@noop {} {\bibfield  {journal} {\bibinfo  {journal} {Nature
				Nanotechnology}\ }\textbf {\bibinfo {volume} {12}},\ \bibinfo {pages} {744}
		(\bibinfo {year} {2017})}\BibitemShut {NoStop}%
	\bibitem [{\citenamefont {Zhang}\ \emph {et~al.}(2017)\citenamefont {Zhang},
		\citenamefont {Jia}, \citenamefont {Kholmanov}, \citenamefont {Dong},
		\citenamefont {Er}, \citenamefont {Chen}, \citenamefont {Guo}, \citenamefont
		{Jin}, \citenamefont {Shenoy}, \citenamefont {Shi} \emph
		{et~al.}}]{zhang2017janus}%
	\BibitemOpen
	\bibfield  {author} {\bibinfo {author} {\bibfnamefont {J.}~\bibnamefont
			{Zhang}}, \bibinfo {author} {\bibfnamefont {S.}~\bibnamefont {Jia}}, \bibinfo
		{author} {\bibfnamefont {I.}~\bibnamefont {Kholmanov}}, \bibinfo {author}
		{\bibfnamefont {L.}~\bibnamefont {Dong}}, \bibinfo {author} {\bibfnamefont
			{D.}~\bibnamefont {Er}}, \bibinfo {author} {\bibfnamefont {W.}~\bibnamefont
			{Chen}}, \bibinfo {author} {\bibfnamefont {H.}~\bibnamefont {Guo}}, \bibinfo
		{author} {\bibfnamefont {Z.}~\bibnamefont {Jin}}, \bibinfo {author}
		{\bibfnamefont {V.~B.}\ \bibnamefont {Shenoy}}, \bibinfo {author}
		{\bibfnamefont {L.}~\bibnamefont {Shi}},  \emph {et~al.},\ }\href@noop {}
	{\bibfield  {journal} {\bibinfo  {journal} {ACS Nano}\ }\textbf {\bibinfo
			{volume} {11}},\ \bibinfo {pages} {8192} (\bibinfo {year}
		{2017})}\BibitemShut {NoStop}%
	\bibitem [{\citenamefont {Zhu}\ \emph {et~al.}(2025)\citenamefont {Zhu},
		\citenamefont {Chen}, \citenamefont {Fan}, \citenamefont {Tang},
		\citenamefont {Zhan}, \citenamefont {Wong}, \citenamefont {Chen},
		\citenamefont {Wan},\ and\ \citenamefont {Chen}}]{zhu2025robust}%
	\BibitemOpen
	\bibfield  {author} {\bibinfo {author} {\bibfnamefont {Q.}~\bibnamefont
			{Zhu}}, \bibinfo {author} {\bibfnamefont {E.}~\bibnamefont {Chen}}, \bibinfo
		{author} {\bibfnamefont {K.}~\bibnamefont {Fan}}, \bibinfo {author}
		{\bibfnamefont {J.}~\bibnamefont {Tang}}, \bibinfo {author} {\bibfnamefont
			{R.}~\bibnamefont {Zhan}}, \bibinfo {author} {\bibfnamefont {K.~S.}\
			\bibnamefont {Wong}}, \bibinfo {author} {\bibfnamefont {Z.}~\bibnamefont
			{Chen}}, \bibinfo {author} {\bibfnamefont {X.}~\bibnamefont {Wan}}, \ and\
		\bibinfo {author} {\bibfnamefont {K.}~\bibnamefont {Chen}},\ }\href@noop {}
	{\bibfield  {journal} {\bibinfo  {journal} {Small Methods}\ }\textbf
		{\bibinfo {volume} {9}},\ \bibinfo {pages} {2401310} (\bibinfo {year}
		{2025})}\BibitemShut {NoStop}%
	\bibitem [{\citenamefont {Xiong}\ \emph {et~al.}(2018)\citenamefont {Xiong},
		\citenamefont {Zhou}, \citenamefont {Zhang}, \citenamefont {Kitamura},\ and\
		\citenamefont {Li}}]{xiong2018spontaneous}%
	\BibitemOpen
	\bibfield  {author} {\bibinfo {author} {\bibfnamefont {Q.-L.}\ \bibnamefont
			{Xiong}}, \bibinfo {author} {\bibfnamefont {J.}~\bibnamefont {Zhou}},
		\bibinfo {author} {\bibfnamefont {J.}~\bibnamefont {Zhang}}, \bibinfo
		{author} {\bibfnamefont {T.}~\bibnamefont {Kitamura}}, \ and\ \bibinfo
		{author} {\bibfnamefont {Z.-H.}\ \bibnamefont {Li}},\ }\href@noop {}
	{\bibfield  {journal} {\bibinfo  {journal} {Physical Chemistry Chemical
				Physics}\ }\textbf {\bibinfo {volume} {20}},\ \bibinfo {pages} {20988}
		(\bibinfo {year} {2018})}\BibitemShut {NoStop}%
	\bibitem [{\citenamefont {Ye}\ \emph {et~al.}(2020)\citenamefont {Ye},
		\citenamefont {Zhang}, \citenamefont {Wei}, \citenamefont {Han},
		\citenamefont {Liu}, \citenamefont {Liu}, \citenamefont {Yin},\ and\
		\citenamefont {Wang}}]{ye2020intrinsic}%
	\BibitemOpen
	\bibfield  {author} {\bibinfo {author} {\bibfnamefont {H.}~\bibnamefont
			{Ye}}, \bibinfo {author} {\bibfnamefont {Y.}~\bibnamefont {Zhang}}, \bibinfo
		{author} {\bibfnamefont {A.}~\bibnamefont {Wei}}, \bibinfo {author}
		{\bibfnamefont {D.}~\bibnamefont {Han}}, \bibinfo {author} {\bibfnamefont
			{Y.}~\bibnamefont {Liu}}, \bibinfo {author} {\bibfnamefont {W.}~\bibnamefont
			{Liu}}, \bibinfo {author} {\bibfnamefont {Y.}~\bibnamefont {Yin}}, \ and\
		\bibinfo {author} {\bibfnamefont {M.}~\bibnamefont {Wang}},\ }\href@noop {}
	{\bibfield  {journal} {\bibinfo  {journal} {Applied Surface Science}\
		}\textbf {\bibinfo {volume} {519}},\ \bibinfo {pages} {146251} (\bibinfo
		{year} {2020})}\BibitemShut {NoStop}%
	\bibitem [{\citenamefont {Yang}\ \emph {et~al.}(2024)\citenamefont {Yang},
		\citenamefont {Ye}, \citenamefont {Sun}, \citenamefont {Wu}, \citenamefont
		{Liu},\ and\ \citenamefont {Liu}}]{yang2024unveiling}%
	\BibitemOpen
	\bibfield  {author} {\bibinfo {author} {\bibfnamefont {R.}~\bibnamefont
			{Yang}}, \bibinfo {author} {\bibfnamefont {H.}~\bibnamefont {Ye}}, \bibinfo
		{author} {\bibfnamefont {N.}~\bibnamefont {Sun}}, \bibinfo {author}
		{\bibfnamefont {Z.}~\bibnamefont {Wu}}, \bibinfo {author} {\bibfnamefont
			{Y.}~\bibnamefont {Liu}}, \ and\ \bibinfo {author} {\bibfnamefont
			{W.}~\bibnamefont {Liu}},\ }\href@noop {} {\bibfield  {journal} {\bibinfo
			{journal} {ACS Applied Materials \& Interfaces}\ }\textbf {\bibinfo {volume}
			{16}},\ \bibinfo {pages} {43860} (\bibinfo {year} {2024})}\BibitemShut
	{NoStop}%
	\bibitem [{\citenamefont {Dong}, \citenamefont {Lou},\ and\ \citenamefont
		{Shenoy}(2017)}]{dong2017large}%
	\BibitemOpen
	\bibfield  {author} {\bibinfo {author} {\bibfnamefont {L.}~\bibnamefont
			{Dong}}, \bibinfo {author} {\bibfnamefont {J.}~\bibnamefont {Lou}}, \ and\
		\bibinfo {author} {\bibfnamefont {V.~B.}\ \bibnamefont {Shenoy}},\
	}\href@noop {} {\bibfield  {journal} {\bibinfo  {journal} {ACS Nano}\
		}\textbf {\bibinfo {volume} {11}},\ \bibinfo {pages} {8242} (\bibinfo {year}
		{2017})}\BibitemShut {NoStop}%
	\bibitem [{\citenamefont {Rawat}\ \emph {et~al.}(2020)\citenamefont {Rawat},
		\citenamefont {Mohanta}, \citenamefont {Jena}, \citenamefont {Dimple},
		\citenamefont {Ahammed},\ and\ \citenamefont
		{De~Sarkar}}]{rawat2020nanoscale}%
	\BibitemOpen
	\bibfield  {author} {\bibinfo {author} {\bibfnamefont {A.}~\bibnamefont
			{Rawat}}, \bibinfo {author} {\bibfnamefont {M.~K.}\ \bibnamefont {Mohanta}},
		\bibinfo {author} {\bibfnamefont {N.}~\bibnamefont {Jena}}, \bibinfo {author}
		{\bibnamefont {Dimple}}, \bibinfo {author} {\bibfnamefont {R.}~\bibnamefont
			{Ahammed}}, \ and\ \bibinfo {author} {\bibfnamefont {A.}~\bibnamefont
			{De~Sarkar}},\ }\href@noop {} {\bibfield  {journal} {\bibinfo  {journal} {The
				Journal of Physical Chemistry C}\ }\textbf {\bibinfo {volume} {124}},\
		\bibinfo {pages} {10385} (\bibinfo {year} {2020})}\BibitemShut {NoStop}%
	\bibitem [{\citenamefont {Cai}\ \emph {et~al.}(2019)\citenamefont {Cai},
		\citenamefont {Guo}, \citenamefont {Gao},\ and\ \citenamefont
		{Guo}}]{cai2019tribo}%
	\BibitemOpen
	\bibfield  {author} {\bibinfo {author} {\bibfnamefont {H.}~\bibnamefont
			{Cai}}, \bibinfo {author} {\bibfnamefont {Y.}~\bibnamefont {Guo}}, \bibinfo
		{author} {\bibfnamefont {H.}~\bibnamefont {Gao}}, \ and\ \bibinfo {author}
		{\bibfnamefont {W.}~\bibnamefont {Guo}},\ }\href@noop {} {\bibfield
		{journal} {\bibinfo  {journal} {Nano Energy}\ }\textbf {\bibinfo {volume}
			{56}},\ \bibinfo {pages} {33} (\bibinfo {year} {2019})}\BibitemShut {NoStop}%
	\bibitem [{\citenamefont {Li}, \citenamefont {Guo},\ and\ \citenamefont
		{Guo}(2022)}]{li2022electro}%
	\BibitemOpen
	\bibfield  {author} {\bibinfo {author} {\bibfnamefont {H.}~\bibnamefont
			{Li}}, \bibinfo {author} {\bibfnamefont {Y.}~\bibnamefont {Guo}}, \ and\
		\bibinfo {author} {\bibfnamefont {W.}~\bibnamefont {Guo}},\ }\href@noop {}
	{\bibfield  {journal} {\bibinfo  {journal} {Friction}\ }\textbf {\bibinfo
			{volume} {10}},\ \bibinfo {pages} {1851} (\bibinfo {year}
		{2022})}\BibitemShut {NoStop}%
	\bibitem [{\citenamefont {Smaili}\ \emph {et~al.}(2021)\citenamefont {Smaili},
		\citenamefont {Laref}, \citenamefont {Garcia}, \citenamefont
		{Schwingenschl{\"o}gl}, \citenamefont {Roche},\ and\ \citenamefont
		{Manchon}}]{smaili2021janus}%
	\BibitemOpen
	\bibfield  {author} {\bibinfo {author} {\bibfnamefont {I.}~\bibnamefont
			{Smaili}}, \bibinfo {author} {\bibfnamefont {S.}~\bibnamefont {Laref}},
		\bibinfo {author} {\bibfnamefont {J.~H.}\ \bibnamefont {Garcia}}, \bibinfo
		{author} {\bibfnamefont {U.}~\bibnamefont {Schwingenschl{\"o}gl}}, \bibinfo
		{author} {\bibfnamefont {S.}~\bibnamefont {Roche}}, \ and\ \bibinfo {author}
		{\bibfnamefont {A.}~\bibnamefont {Manchon}},\ }\href@noop {} {\bibfield
		{journal} {\bibinfo  {journal} {Physical Review B}\ }\textbf {\bibinfo
			{volume} {104}},\ \bibinfo {pages} {104415} (\bibinfo {year}
		{2021})}\BibitemShut {NoStop}%
	\bibitem [{\citenamefont {Yu}\ \emph {et~al.}(2021)\citenamefont {Yu},
		\citenamefont {Zhou}, \citenamefont {Zhang},\ and\ \citenamefont
		{Chang}}]{yu2021spin}%
	\BibitemOpen
	\bibfield  {author} {\bibinfo {author} {\bibfnamefont {S.-B.}\ \bibnamefont
			{Yu}}, \bibinfo {author} {\bibfnamefont {M.}~\bibnamefont {Zhou}}, \bibinfo
		{author} {\bibfnamefont {D.}~\bibnamefont {Zhang}}, \ and\ \bibinfo {author}
		{\bibfnamefont {K.}~\bibnamefont {Chang}},\ }\href@noop {} {\bibfield
		{journal} {\bibinfo  {journal} {Physical Review B}\ }\textbf {\bibinfo
			{volume} {104}},\ \bibinfo {pages} {075435} (\bibinfo {year}
		{2021})}\BibitemShut {NoStop}%
	\bibitem [{\citenamefont {Lin}\ \emph {et~al.}(2025)\citenamefont {Lin},
		\citenamefont {Rouzhahong}, \citenamefont {Liang}, \citenamefont {Yuan},
		\citenamefont {Yao},\ and\ \citenamefont {Li}}]{lin2025rashba}%
	\BibitemOpen
	\bibfield  {author} {\bibinfo {author} {\bibfnamefont {X.}~\bibnamefont
			{Lin}}, \bibinfo {author} {\bibfnamefont {Y.}~\bibnamefont {Rouzhahong}},
		\bibinfo {author} {\bibfnamefont {C.}~\bibnamefont {Liang}}, \bibinfo
		{author} {\bibfnamefont {J.}~\bibnamefont {Yuan}}, \bibinfo {author}
		{\bibfnamefont {S.}~\bibnamefont {Yao}}, \ and\ \bibinfo {author}
		{\bibfnamefont {H.}~\bibnamefont {Li}},\ }\href@noop {} {\bibfield  {journal}
		{\bibinfo  {journal} {Physical Review B}\ }\textbf {\bibinfo {volume}
			{111}},\ \bibinfo {pages} {195432} (\bibinfo {year} {2025})}\BibitemShut
	{NoStop}%
	\bibitem [{\citenamefont {Li}\ \emph {et~al.}(2025)\citenamefont {Li},
		\citenamefont {Lian}, \citenamefont {Yang}, \citenamefont {Huang},
		\citenamefont {Xu}, \citenamefont {Wang}, \citenamefont {Huang},
		\citenamefont {Hu}, \citenamefont {Huang},\ and\ \citenamefont
		{Duan}}]{li2025key}%
	\BibitemOpen
	\bibfield  {author} {\bibinfo {author} {\bibfnamefont {L.}~\bibnamefont
			{Li}}, \bibinfo {author} {\bibfnamefont {J.-C.}\ \bibnamefont {Lian}},
		\bibinfo {author} {\bibfnamefont {Z.-X.}\ \bibnamefont {Yang}}, \bibinfo
		{author} {\bibfnamefont {T.}~\bibnamefont {Huang}}, \bibinfo {author}
		{\bibfnamefont {J.-Q.}\ \bibnamefont {Xu}}, \bibinfo {author} {\bibfnamefont
			{X.}~\bibnamefont {Wang}}, \bibinfo {author} {\bibfnamefont {G.-F.}\
			\bibnamefont {Huang}}, \bibinfo {author} {\bibfnamefont {W.}~\bibnamefont
			{Hu}}, \bibinfo {author} {\bibfnamefont {W.-Q.}\ \bibnamefont {Huang}}, \
		and\ \bibinfo {author} {\bibfnamefont {X.}~\bibnamefont {Duan}},\ }\href@noop
	{} {\bibfield  {journal} {\bibinfo  {journal} {Physical Review B}\ }\textbf
		{\bibinfo {volume} {112}},\ \bibinfo {pages} {075423} (\bibinfo {year}
		{2025})}\BibitemShut {NoStop}%
	\bibitem [{\citenamefont {Er}\ \emph {et~al.}(2018)\citenamefont {Er},
		\citenamefont {Ye}, \citenamefont {Frey}, \citenamefont {Kumar},
		\citenamefont {Lou},\ and\ \citenamefont {Shenoy}}]{er2018prediction}%
	\BibitemOpen
	\bibfield  {author} {\bibinfo {author} {\bibfnamefont {D.}~\bibnamefont
			{Er}}, \bibinfo {author} {\bibfnamefont {H.}~\bibnamefont {Ye}}, \bibinfo
		{author} {\bibfnamefont {N.~C.}\ \bibnamefont {Frey}}, \bibinfo {author}
		{\bibfnamefont {H.}~\bibnamefont {Kumar}}, \bibinfo {author} {\bibfnamefont
			{J.}~\bibnamefont {Lou}}, \ and\ \bibinfo {author} {\bibfnamefont {V.~B.}\
			\bibnamefont {Shenoy}},\ }\href@noop {} {\bibfield  {journal} {\bibinfo
			{journal} {Nano Letters}\ }\textbf {\bibinfo {volume} {18}},\ \bibinfo
		{pages} {3943} (\bibinfo {year} {2018})}\BibitemShut {NoStop}%
	\bibitem [{\citenamefont {Popov}(2004)}]{popov2004carbon}%
	\BibitemOpen
	\bibfield  {author} {\bibinfo {author} {\bibfnamefont {V.~N.}\ \bibnamefont
			{Popov}},\ }\href@noop {} {\bibfield  {journal} {\bibinfo  {journal}
			{Materials Science and Engineering: R: Reports}\ }\textbf {\bibinfo {volume}
			{43}},\ \bibinfo {pages} {61} (\bibinfo {year} {2004})}\BibitemShut {NoStop}%
	\bibitem [{\citenamefont {Popov}\ and\ \citenamefont
		{Lambin}(2006)}]{popov2006radius}%
	\BibitemOpen
	\bibfield  {author} {\bibinfo {author} {\bibfnamefont {V.~N.}\ \bibnamefont
			{Popov}}\ and\ \bibinfo {author} {\bibfnamefont {P.}~\bibnamefont {Lambin}},\
	}\href@noop {} {\bibfield  {journal} {\bibinfo  {journal} {Physical Review
				B}\ }\textbf {\bibinfo {volume} {73}},\ \bibinfo {pages} {085407} (\bibinfo
		{year} {2006})}\BibitemShut {NoStop}%
	\bibitem [{\citenamefont {Xiang}\ \emph {et~al.}(2020)\citenamefont {Xiang},
		\citenamefont {Inoue}, \citenamefont {Zheng}, \citenamefont {Kumamoto},
		\citenamefont {Qian}, \citenamefont {Sato}, \citenamefont {Liu},
		\citenamefont {Tang}, \citenamefont {Gokhale}, \citenamefont {Guo},
		\citenamefont {Hisama}, \citenamefont {Yotsumoto}, \citenamefont {Ogamoto},
		\citenamefont {Arai}, \citenamefont {Kobayashi}, \citenamefont {Zhang},
		\citenamefont {Hou}, \citenamefont {Anisimov}, \citenamefont {Maruyama},
		\citenamefont {Miyata}, \citenamefont {Okada}, \citenamefont {Chiashi},
		\citenamefont {Li}, \citenamefont {Kong}, \citenamefont {Kauppinen},
		\citenamefont {Ikuhara}, \citenamefont {Suenaga},\ and\ \citenamefont
		{Maruyama}}]{XiangR2020sci}%
	\BibitemOpen
	\bibfield  {author} {\bibinfo {author} {\bibfnamefont {R.}~\bibnamefont
			{Xiang}}, \bibinfo {author} {\bibfnamefont {T.}~\bibnamefont {Inoue}},
		\bibinfo {author} {\bibfnamefont {Y.}~\bibnamefont {Zheng}}, \bibinfo
		{author} {\bibfnamefont {A.}~\bibnamefont {Kumamoto}}, \bibinfo {author}
		{\bibfnamefont {Y.}~\bibnamefont {Qian}}, \bibinfo {author} {\bibfnamefont
			{Y.}~\bibnamefont {Sato}}, \bibinfo {author} {\bibfnamefont {M.}~\bibnamefont
			{Liu}}, \bibinfo {author} {\bibfnamefont {D.}~\bibnamefont {Tang}}, \bibinfo
		{author} {\bibfnamefont {D.}~\bibnamefont {Gokhale}}, \bibinfo {author}
		{\bibfnamefont {J.}~\bibnamefont {Guo}}, \bibinfo {author} {\bibfnamefont
			{K.}~\bibnamefont {Hisama}}, \bibinfo {author} {\bibfnamefont
			{S.}~\bibnamefont {Yotsumoto}}, \bibinfo {author} {\bibfnamefont
			{T.}~\bibnamefont {Ogamoto}}, \bibinfo {author} {\bibfnamefont
			{H.}~\bibnamefont {Arai}}, \bibinfo {author} {\bibfnamefont {Y.}~\bibnamefont
			{Kobayashi}}, \bibinfo {author} {\bibfnamefont {H.}~\bibnamefont {Zhang}},
		\bibinfo {author} {\bibfnamefont {B.}~\bibnamefont {Hou}}, \bibinfo {author}
		{\bibfnamefont {A.}~\bibnamefont {Anisimov}}, \bibinfo {author}
		{\bibfnamefont {M.}~\bibnamefont {Maruyama}}, \bibinfo {author}
		{\bibfnamefont {Y.}~\bibnamefont {Miyata}}, \bibinfo {author} {\bibfnamefont
			{S.}~\bibnamefont {Okada}}, \bibinfo {author} {\bibfnamefont
			{S.}~\bibnamefont {Chiashi}}, \bibinfo {author} {\bibfnamefont
			{Y.}~\bibnamefont {Li}}, \bibinfo {author} {\bibfnamefont {J.}~\bibnamefont
			{Kong}}, \bibinfo {author} {\bibfnamefont {E.~I.}\ \bibnamefont {Kauppinen}},
		\bibinfo {author} {\bibfnamefont {Y.}~\bibnamefont {Ikuhara}}, \bibinfo
		{author} {\bibfnamefont {K.}~\bibnamefont {Suenaga}}, \ and\ \bibinfo
		{author} {\bibfnamefont {S.}~\bibnamefont {Maruyama}},\ }\href@noop {}
	{\bibfield  {journal} {\bibinfo  {journal} {Science}\ }\textbf {\bibinfo
			{volume} {367}},\ \bibinfo {pages} {537} (\bibinfo {year}
		{2020})}\BibitemShut {NoStop}%
	\bibitem [{\citenamefont {Wang}\ \emph {et~al.}(2020)\citenamefont {Wang},
		\citenamefont {Zheng}, \citenamefont {Inoue}, \citenamefont {Xiang},
		\citenamefont {Shawky}, \citenamefont {Watanabe}, \citenamefont {Anisimov},
		\citenamefont {Kauppinen}, \citenamefont {Chiashi},\ and\ \citenamefont
		{Maruyama}}]{WangP2020acsn}%
	\BibitemOpen
	\bibfield  {author} {\bibinfo {author} {\bibfnamefont {P.}~\bibnamefont
			{Wang}}, \bibinfo {author} {\bibfnamefont {Y.}~\bibnamefont {Zheng}},
		\bibinfo {author} {\bibfnamefont {T.}~\bibnamefont {Inoue}}, \bibinfo
		{author} {\bibfnamefont {R.}~\bibnamefont {Xiang}}, \bibinfo {author}
		{\bibfnamefont {A.}~\bibnamefont {Shawky}}, \bibinfo {author} {\bibfnamefont
			{M.}~\bibnamefont {Watanabe}}, \bibinfo {author} {\bibfnamefont
			{A.}~\bibnamefont {Anisimov}}, \bibinfo {author} {\bibfnamefont {E.~I.}\
			\bibnamefont {Kauppinen}}, \bibinfo {author} {\bibfnamefont {S.}~\bibnamefont
			{Chiashi}}, \ and\ \bibinfo {author} {\bibfnamefont {S.}~\bibnamefont
			{Maruyama}},\ }\href@noop {} {\bibfield  {journal} {\bibinfo  {journal} {ACS
				Nano}\ }\textbf {\bibinfo {volume} {14}},\ \bibinfo {pages} {4298} (\bibinfo
		{year} {2020})}\BibitemShut {NoStop}%
	\bibitem [{\citenamefont {Cambr{\'e}}\ \emph {et~al.}(2021)\citenamefont
		{Cambr{\'e}}, \citenamefont {Liu}, \citenamefont {Levshov}, \citenamefont
		{Otsuka}, \citenamefont {Maruyama},\ and\ \citenamefont
		{Xiang}}]{CambreS2021small}%
	\BibitemOpen
	\bibfield  {author} {\bibinfo {author} {\bibfnamefont {S.}~\bibnamefont
			{Cambr{\'e}}}, \bibinfo {author} {\bibfnamefont {M.}~\bibnamefont {Liu}},
		\bibinfo {author} {\bibfnamefont {D.}~\bibnamefont {Levshov}}, \bibinfo
		{author} {\bibfnamefont {K.}~\bibnamefont {Otsuka}}, \bibinfo {author}
		{\bibfnamefont {S.}~\bibnamefont {Maruyama}}, \ and\ \bibinfo {author}
		{\bibfnamefont {R.}~\bibnamefont {Xiang}},\ }\href@noop {} {\bibfield
		{journal} {\bibinfo  {journal} {Small}\ }\textbf {\bibinfo {volume} {17}},\
		\bibinfo {pages} {2170196} (\bibinfo {year} {2021})}\BibitemShut {NoStop}%
	\bibitem [{\citenamefont {Tang}\ \emph {et~al.}(2024)\citenamefont {Tang},
		\citenamefont {Cretu}, \citenamefont {Ishihara}, \citenamefont {Zheng},
		\citenamefont {Otsuka}, \citenamefont {Xiang}, \citenamefont {Maruyama},
		\citenamefont {Cheng}, \citenamefont {Liu},\ and\ \citenamefont
		{Golberg}}]{tang2024chirality}%
	\BibitemOpen
	\bibfield  {author} {\bibinfo {author} {\bibfnamefont {D.-M.}\ \bibnamefont
			{Tang}}, \bibinfo {author} {\bibfnamefont {O.}~\bibnamefont {Cretu}},
		\bibinfo {author} {\bibfnamefont {S.}~\bibnamefont {Ishihara}}, \bibinfo
		{author} {\bibfnamefont {Y.}~\bibnamefont {Zheng}}, \bibinfo {author}
		{\bibfnamefont {K.}~\bibnamefont {Otsuka}}, \bibinfo {author} {\bibfnamefont
			{R.}~\bibnamefont {Xiang}}, \bibinfo {author} {\bibfnamefont
			{S.}~\bibnamefont {Maruyama}}, \bibinfo {author} {\bibfnamefont {H.-M.}\
			\bibnamefont {Cheng}}, \bibinfo {author} {\bibfnamefont {C.}~\bibnamefont
			{Liu}}, \ and\ \bibinfo {author} {\bibfnamefont {D.}~\bibnamefont
			{Golberg}},\ }\href@noop {} {\bibfield  {journal} {\bibinfo  {journal}
			{Nature Reviews Electrical Engineering}\ }\textbf {\bibinfo {volume} {1}},\
		\bibinfo {pages} {149} (\bibinfo {year} {2024})}\BibitemShut {NoStop}%
	\bibitem [{\citenamefont {Plimpton}(1995)}]{PlimptonSJ}%
	\BibitemOpen
	\bibfield  {author} {\bibinfo {author} {\bibfnamefont {S.~J.}\ \bibnamefont
			{Plimpton}},\ }\href@noop {} {\bibfield  {journal} {\bibinfo  {journal} {J.
				Comput. Phys.}\ }\textbf {\bibinfo {volume} {117}},\ \bibinfo {pages} {1}
		(\bibinfo {year} {1995})}\BibitemShut {NoStop}%
	\bibitem [{\citenamefont {Stukowski}(2010)}]{ovito}%
	\BibitemOpen
	\bibfield  {author} {\bibinfo {author} {\bibfnamefont {A.}~\bibnamefont
			{Stukowski}},\ }\href@noop {} {\bibfield  {journal} {\bibinfo  {journal}
			{Modelling Simul. Mater. Sci. Eng.}\ }\textbf {\bibinfo {volume} {18}},\
		\bibinfo {pages} {015012} (\bibinfo {year} {2010})}\BibitemShut {NoStop}%
	\bibitem [{\citenamefont {Stillinger}\ and\ \citenamefont
		{Weber}(1985)}]{StillingerFH}%
	\BibitemOpen
	\bibfield  {author} {\bibinfo {author} {\bibfnamefont {F.~H.}\ \bibnamefont
			{Stillinger}}\ and\ \bibinfo {author} {\bibfnamefont {T.~A.}\ \bibnamefont
			{Weber}},\ }\href@noop {} {\bibfield  {journal} {\bibinfo  {journal} {Phys.
				Rev. B}\ }\textbf {\bibinfo {volume} {31}},\ \bibinfo {pages} {5262}
		(\bibinfo {year} {1985})}\BibitemShut {NoStop}%
	\bibitem [{\citenamefont {Jiang}(2019)}]{JiangJW2018swmx2}%
	\BibitemOpen
	\bibfield  {author} {\bibinfo {author} {\bibfnamefont {J.-W.}\ \bibnamefont
			{Jiang}},\ }\href@noop {} {\bibfield  {journal} {\bibinfo  {journal} {Acta
				Mechanica Solida Sinica}\ }\textbf {\bibinfo {volume} {32}},\ \bibinfo
		{pages} {17} (\bibinfo {year} {2019})}\BibitemShut {NoStop}%
	\bibitem [{\citenamefont {Perdew}, \citenamefont {Burke},\ and\ \citenamefont
		{Ernzerhof}(1996)}]{PerdewJP1996prl}%
	\BibitemOpen
	\bibfield  {author} {\bibinfo {author} {\bibfnamefont {J.~P.}\ \bibnamefont
			{Perdew}}, \bibinfo {author} {\bibfnamefont {K.}~\bibnamefont {Burke}}, \
		and\ \bibinfo {author} {\bibfnamefont {M.}~\bibnamefont {Ernzerhof}},\
	}\href@noop {} {\bibfield  {journal} {\bibinfo  {journal} {Phys. Rev. Lett.}\
		}\textbf {\bibinfo {volume} {77}},\ \bibinfo {pages} {3865} (\bibinfo {year}
		{1996})}\BibitemShut {NoStop}%
	\bibitem [{\citenamefont {Gale}(1997)}]{gulp}%
	\BibitemOpen
	\bibfield  {author} {\bibinfo {author} {\bibfnamefont {J.~D.}\ \bibnamefont
			{Gale}},\ }\href@noop {} {\bibfield  {journal} {\bibinfo  {journal} {J. Chem.
				Soc., Faraday Trans.}\ }\textbf {\bibinfo {volume} {93}},\ \bibinfo {pages}
		{629} (\bibinfo {year} {1997})}\BibitemShut {NoStop}%
	\bibitem [{\citenamefont {Arroyo~Balaguer}\ and\ \citenamefont
		{Belytschko}(2004)}]{arroyo2004finite}%
	\BibitemOpen
	\bibfield  {author} {\bibinfo {author} {\bibfnamefont {M.}~\bibnamefont
			{Arroyo~Balaguer}}\ and\ \bibinfo {author} {\bibfnamefont {T.}~\bibnamefont
			{Belytschko}},\ }\href@noop {} {\bibfield  {journal} {\bibinfo  {journal}
			{Physical Review B}\ }\textbf {\bibinfo {volume} {69}},\ \bibinfo {pages}
		{115415} (\bibinfo {year} {2004})}\BibitemShut {NoStop}%
	\bibitem [{\citenamefont {Jiang}\ \emph {et~al.}(2013)\citenamefont {Jiang},
		\citenamefont {Qi}, \citenamefont {Park},\ and\ \citenamefont
		{Rabczuk}}]{jiang2013elastic}%
	\BibitemOpen
	\bibfield  {author} {\bibinfo {author} {\bibfnamefont {J.-W.}\ \bibnamefont
			{Jiang}}, \bibinfo {author} {\bibfnamefont {Z.}~\bibnamefont {Qi}}, \bibinfo
		{author} {\bibfnamefont {H.~S.}\ \bibnamefont {Park}}, \ and\ \bibinfo
		{author} {\bibfnamefont {T.}~\bibnamefont {Rabczuk}},\ }\href@noop {}
	{\bibfield  {journal} {\bibinfo  {journal} {Nanotechnology}\ }\textbf
		{\bibinfo {volume} {24}},\ \bibinfo {pages} {435705} (\bibinfo {year}
		{2013})}\BibitemShut {NoStop}%
	\bibitem [{\citenamefont {Jiang}\ and\ \citenamefont
		{Zhou}(2017)}]{jiang2017handbook}%
	\BibitemOpen
	\bibfield  {author} {\bibinfo {author} {\bibfnamefont {J.-W.}\ \bibnamefont
			{Jiang}}\ and\ \bibinfo {author} {\bibfnamefont {Y.-P.}\ \bibnamefont
			{Zhou}},\ }\href@noop {} {\emph {\bibinfo {title} {Handbook of
				Stillinger-Weber potential parameters for two-dimensional atomic crystals}}}\
	(\bibinfo  {publisher} {BoD--Books on Demand},\ \bibinfo {year}
	{2017})\BibitemShut {NoStop}%
	\bibitem [{\citenamefont {Popov}, \citenamefont {Van~Doren},\ and\
		\citenamefont {Balkanski}(1999)}]{popov1999lattice}%
	\BibitemOpen
	\bibfield  {author} {\bibinfo {author} {\bibfnamefont {V.}~\bibnamefont
			{Popov}}, \bibinfo {author} {\bibfnamefont {V.}~\bibnamefont {Van~Doren}}, \
		and\ \bibinfo {author} {\bibfnamefont {M.}~\bibnamefont {Balkanski}},\
	}\href@noop {} {\bibfield  {journal} {\bibinfo  {journal} {Physical Review
				B}\ }\textbf {\bibinfo {volume} {59}},\ \bibinfo {pages} {8355} (\bibinfo
		{year} {1999})}\BibitemShut {NoStop}%
	\bibitem [{\citenamefont {Jiang}, \citenamefont {Wang},\ and\ \citenamefont
		{Rabczuk}(2014)}]{jiang2014phonon}%
	\BibitemOpen
	\bibfield  {author} {\bibinfo {author} {\bibfnamefont {J.-W.}\ \bibnamefont
			{Jiang}}, \bibinfo {author} {\bibfnamefont {B.-S.}\ \bibnamefont {Wang}}, \
		and\ \bibinfo {author} {\bibfnamefont {T.}~\bibnamefont {Rabczuk}},\
	}\href@noop {} {\bibfield  {journal} {\bibinfo  {journal} {Nanotechnology}\
		}\textbf {\bibinfo {volume} {25}},\ \bibinfo {pages} {105706} (\bibinfo
		{year} {2014})}\BibitemShut {NoStop}%
	\bibitem [{\citenamefont {Damnjanovi{\'c}}\ \emph {et~al.}(1999)\citenamefont
		{Damnjanovi{\'c}}, \citenamefont {Milo{\v{s}}evi{\'c}}, \citenamefont
		{Vukovi{\'c}},\ and\ \citenamefont {Sredanovi{\'c}}}]{damnjanovic1999full}%
	\BibitemOpen
	\bibfield  {author} {\bibinfo {author} {\bibfnamefont {M.}~\bibnamefont
			{Damnjanovi{\'c}}}, \bibinfo {author} {\bibfnamefont {I.}~\bibnamefont
			{Milo{\v{s}}evi{\'c}}}, \bibinfo {author} {\bibfnamefont {T.}~\bibnamefont
			{Vukovi{\'c}}}, \ and\ \bibinfo {author} {\bibfnamefont {R.}~\bibnamefont
			{Sredanovi{\'c}}},\ }\href@noop {} {\bibfield  {journal} {\bibinfo  {journal}
			{Physical Review B}\ }\textbf {\bibinfo {volume} {60}},\ \bibinfo {pages}
		{2728} (\bibinfo {year} {1999})}\BibitemShut {NoStop}%
	\bibitem [{\citenamefont {Damnjanovic}, \citenamefont {Vukovic},\ and\
		\citenamefont {Milosevic}(2000)}]{damnjanovic2000modified}%
	\BibitemOpen
	\bibfield  {author} {\bibinfo {author} {\bibfnamefont {M.}~\bibnamefont
			{Damnjanovic}}, \bibinfo {author} {\bibfnamefont {T.}~\bibnamefont
			{Vukovic}}, \ and\ \bibinfo {author} {\bibfnamefont {I.}~\bibnamefont
			{Milosevic}},\ }\href@noop {} {\bibfield  {journal} {\bibinfo  {journal}
			{Journal of Physics A: Mathematical and General}\ }\textbf {\bibinfo {volume}
			{33}},\ \bibinfo {pages} {6561} (\bibinfo {year} {2000})}\BibitemShut
	{NoStop}%
	\bibitem [{\citenamefont {Tang}, \citenamefont {Wang},\ and\ \citenamefont
		{Su}(2011)}]{tang2011symmetry}%
	\BibitemOpen
	\bibfield  {author} {\bibinfo {author} {\bibfnamefont {H.}~\bibnamefont
			{Tang}}, \bibinfo {author} {\bibfnamefont {B.-S.}\ \bibnamefont {Wang}}, \
		and\ \bibinfo {author} {\bibfnamefont {Z.-B.}\ \bibnamefont {Su}},\
	}\href@noop {} {\bibfield  {journal} {\bibinfo  {journal} {Graphene
				Simulation}\ } (\bibinfo {year} {2011})}\BibitemShut {NoStop}%
	\bibitem [{\citenamefont {Popov}(2003)}]{popov2003lattice}%
	\BibitemOpen
	\bibfield  {author} {\bibinfo {author} {\bibfnamefont {V.~N.}\ \bibnamefont
			{Popov}},\ }\href@noop {} {\bibfield  {journal} {\bibinfo  {journal}
			{Physical Review B}\ }\textbf {\bibinfo {volume} {67}},\ \bibinfo {pages}
		{085408} (\bibinfo {year} {2003})}\BibitemShut {NoStop}%
	\bibitem [{\citenamefont {Jiang}\ \emph {et~al.}(2006)\citenamefont {Jiang},
		\citenamefont {Tang}, \citenamefont {Wang},\ and\ \citenamefont
		{Su}}]{JiangJW2006}%
	\BibitemOpen
	\bibfield  {author} {\bibinfo {author} {\bibfnamefont {J.-W.}\ \bibnamefont
			{Jiang}}, \bibinfo {author} {\bibfnamefont {H.}~\bibnamefont {Tang}},
		\bibinfo {author} {\bibfnamefont {B.-S.}\ \bibnamefont {Wang}}, \ and\
		\bibinfo {author} {\bibfnamefont {Z.-B.}\ \bibnamefont {Su}},\ }\href@noop {}
	{\bibfield  {journal} {\bibinfo  {journal} {Phys. Rev. B}\ }\textbf {\bibinfo
			{volume} {73}},\ \bibinfo {pages} {235434} (\bibinfo {year}
		{2006})}\BibitemShut {NoStop}%
	\bibitem [{\citenamefont {K{\"u}rti}, \citenamefont {Kresse},\ and\
		\citenamefont {Kuzmany}(1998)}]{kurti1998first}%
	\BibitemOpen
	\bibfield  {author} {\bibinfo {author} {\bibfnamefont {J.}~\bibnamefont
			{K{\"u}rti}}, \bibinfo {author} {\bibfnamefont {G.}~\bibnamefont {Kresse}}, \
		and\ \bibinfo {author} {\bibfnamefont {H.}~\bibnamefont {Kuzmany}},\
	}\href@noop {} {\bibfield  {journal} {\bibinfo  {journal} {Physical Review
				B}\ }\textbf {\bibinfo {volume} {58}},\ \bibinfo {pages} {R8869} (\bibinfo
		{year} {1998})}\BibitemShut {NoStop}%
	\bibitem [{\citenamefont {Born}\ and\ \citenamefont {Huang}(1954)}]{BornM}%
	\BibitemOpen
	\bibfield  {author} {\bibinfo {author} {\bibfnamefont {M.}~\bibnamefont
			{Born}}\ and\ \bibinfo {author} {\bibfnamefont {K.}~\bibnamefont {Huang}},\
	}\href@noop {} {\emph {\bibinfo {title} {Dynamical Theory of Crystal
				Lattices}}}\ (\bibinfo  {publisher} {Oxford University Press, Oxford},\
	\bibinfo {year} {1954})\BibitemShut {NoStop}%
	\bibitem [{\citenamefont {Landau}\ and\ \citenamefont
		{Lifshitz}(1995)}]{LandauLD}%
	\BibitemOpen
	\bibfield  {author} {\bibinfo {author} {\bibfnamefont {L.~D.}\ \bibnamefont
			{Landau}}\ and\ \bibinfo {author} {\bibfnamefont {E.~M.}\ \bibnamefont
			{Lifshitz}},\ }\href@noop {} {\emph {\bibinfo {title} {Theory of
				Elasticity}}}\ (\bibinfo  {publisher} {Pergamon,Oxford},\ \bibinfo {year}
	{1995})\BibitemShut {NoStop}%
	\bibitem [{\citenamefont {Scott}(1974)}]{scott1974soft}%
	\BibitemOpen
	\bibfield  {author} {\bibinfo {author} {\bibfnamefont {J.}~\bibnamefont
			{Scott}},\ }\href@noop {} {\bibfield  {journal} {\bibinfo  {journal} {Reviews
				of Modern Physics}\ }\textbf {\bibinfo {volume} {46}},\ \bibinfo {pages} {83}
		(\bibinfo {year} {1974})}\BibitemShut {NoStop}%
	\bibitem [{\citenamefont {Nakanishi}, \citenamefont {Nagasawa},\ and\
		\citenamefont {Murakami}(1982)}]{nakanishi1982lattice}%
	\BibitemOpen
	\bibfield  {author} {\bibinfo {author} {\bibfnamefont {N.}~\bibnamefont
			{Nakanishi}}, \bibinfo {author} {\bibfnamefont {A.}~\bibnamefont {Nagasawa}},
		\ and\ \bibinfo {author} {\bibfnamefont {Y.}~\bibnamefont {Murakami}},\
	}\href@noop {} {\bibfield  {journal} {\bibinfo  {journal} {Le Journal de
				Physique Colloques}\ }\textbf {\bibinfo {volume} {43}},\ \bibinfo {pages}
		{C4} (\bibinfo {year} {1982})}\BibitemShut {NoStop}%
	\bibitem [{\citenamefont {Rudin}(2018)}]{rudin2018generalization}%
	\BibitemOpen
	\bibfield  {author} {\bibinfo {author} {\bibfnamefont {S.~P.}\ \bibnamefont
			{Rudin}},\ }\href@noop {} {\bibfield  {journal} {\bibinfo  {journal}
			{Physical Review B}\ }\textbf {\bibinfo {volume} {97}},\ \bibinfo {pages}
		{134114} (\bibinfo {year} {2018})}\BibitemShut {NoStop}%
	\bibitem [{\citenamefont {Gupta}\ \emph {et~al.}(2022)\citenamefont {Gupta},
		\citenamefont {Kumar}, \citenamefont {Mittal},\ and\ \citenamefont
		{Chaplot}}]{gupta2022soft}%
	\BibitemOpen
	\bibfield  {author} {\bibinfo {author} {\bibfnamefont {M.~K.}\ \bibnamefont
			{Gupta}}, \bibinfo {author} {\bibfnamefont {S.}~\bibnamefont {Kumar}},
		\bibinfo {author} {\bibfnamefont {R.}~\bibnamefont {Mittal}}, \ and\ \bibinfo
		{author} {\bibfnamefont {S.~L.}\ \bibnamefont {Chaplot}},\ }\href@noop {}
	{\bibfield  {journal} {\bibinfo  {journal} {Physical Review B}\ }\textbf
		{\bibinfo {volume} {106}},\ \bibinfo {pages} {014311} (\bibinfo {year}
		{2022})}\BibitemShut {NoStop}%
	\bibitem [{\citenamefont {Pallikara}\ \emph {et~al.}(2022)\citenamefont
		{Pallikara}, \citenamefont {Kayastha}, \citenamefont {Skelton},\ and\
		\citenamefont {Whalley}}]{pallikara2022physical}%
	\BibitemOpen
	\bibfield  {author} {\bibinfo {author} {\bibfnamefont {I.}~\bibnamefont
			{Pallikara}}, \bibinfo {author} {\bibfnamefont {P.}~\bibnamefont {Kayastha}},
		\bibinfo {author} {\bibfnamefont {J.~M.}\ \bibnamefont {Skelton}}, \ and\
		\bibinfo {author} {\bibfnamefont {L.~D.}\ \bibnamefont {Whalley}},\
	}\href@noop {} {\bibfield  {journal} {\bibinfo  {journal} {Electronic
				Structure}\ }\textbf {\bibinfo {volume} {4}},\ \bibinfo {pages} {033002}
		(\bibinfo {year} {2022})}\BibitemShut {NoStop}%
	\bibitem [{\citenamefont {Parlinski}, \citenamefont {Li},\ and\ \citenamefont
		{Kawazoe}(1997)}]{parlinski1997first}%
	\BibitemOpen
	\bibfield  {author} {\bibinfo {author} {\bibfnamefont {K.}~\bibnamefont
			{Parlinski}}, \bibinfo {author} {\bibfnamefont {Z.}~\bibnamefont {Li}}, \
		and\ \bibinfo {author} {\bibfnamefont {Y.}~\bibnamefont {Kawazoe}},\
	}\href@noop {} {\bibfield  {journal} {\bibinfo  {journal} {Physical Review
				Letters}\ }\textbf {\bibinfo {volume} {78}},\ \bibinfo {pages} {4063}
		(\bibinfo {year} {1997})}\BibitemShut {NoStop}%
	\bibitem [{\citenamefont {Baroni}\ \emph {et~al.}(2001)\citenamefont {Baroni},
		\citenamefont {De~Gironcoli}, \citenamefont {Dal~Corso},\ and\ \citenamefont
		{Giannozzi}}]{baroni2001phonons}%
	\BibitemOpen
	\bibfield  {author} {\bibinfo {author} {\bibfnamefont {S.}~\bibnamefont
			{Baroni}}, \bibinfo {author} {\bibfnamefont {S.}~\bibnamefont
			{De~Gironcoli}}, \bibinfo {author} {\bibfnamefont {A.}~\bibnamefont
			{Dal~Corso}}, \ and\ \bibinfo {author} {\bibfnamefont {P.}~\bibnamefont
			{Giannozzi}},\ }\href@noop {} {\bibfield  {journal} {\bibinfo  {journal}
			{Reviews of Modern Physics}\ }\textbf {\bibinfo {volume} {73}},\ \bibinfo
		{pages} {515} (\bibinfo {year} {2001})}\BibitemShut {NoStop}%
	\bibitem [{\citenamefont {Togo}, \citenamefont {Oba},\ and\ \citenamefont
		{Tanaka}(2008)}]{togo2008first}%
	\BibitemOpen
	\bibfield  {author} {\bibinfo {author} {\bibfnamefont {A.}~\bibnamefont
			{Togo}}, \bibinfo {author} {\bibfnamefont {F.}~\bibnamefont {Oba}}, \ and\
		\bibinfo {author} {\bibfnamefont {I.}~\bibnamefont {Tanaka}},\ }\href@noop {}
	{\bibfield  {journal} {\bibinfo  {journal} {Physical Review B}\ }\textbf
		{\bibinfo {volume} {78}},\ \bibinfo {pages} {134106} (\bibinfo {year}
		{2008})}\BibitemShut {NoStop}%
	\bibitem [{\citenamefont {Cochran}(1960)}]{cochran1960crystal}%
	\BibitemOpen
	\bibfield  {author} {\bibinfo {author} {\bibfnamefont {W.}~\bibnamefont
			{Cochran}},\ }\href@noop {} {\bibfield  {journal} {\bibinfo  {journal}
			{Advances in Physics}\ }\textbf {\bibinfo {volume} {9}},\ \bibinfo {pages}
		{387} (\bibinfo {year} {1960})}\BibitemShut {NoStop}%
	\bibitem [{\citenamefont {Cowley}(1964)}]{cowley1964lattice}%
	\BibitemOpen
	\bibfield  {author} {\bibinfo {author} {\bibfnamefont {R.}~\bibnamefont
			{Cowley}},\ }\href@noop {} {\bibfield  {journal} {\bibinfo  {journal}
			{Physical Review}\ }\textbf {\bibinfo {volume} {134}},\ \bibinfo {pages}
		{A981} (\bibinfo {year} {1964})}\BibitemShut {NoStop}%
	\bibitem [{\citenamefont {Zhong}, \citenamefont {Vanderbilt},\ and\
		\citenamefont {Rabe}(1995)}]{zhong1995first}%
	\BibitemOpen
	\bibfield  {author} {\bibinfo {author} {\bibfnamefont {W.}~\bibnamefont
			{Zhong}}, \bibinfo {author} {\bibfnamefont {D.}~\bibnamefont {Vanderbilt}}, \
		and\ \bibinfo {author} {\bibfnamefont {K.}~\bibnamefont {Rabe}},\ }\href@noop
	{} {\bibfield  {journal} {\bibinfo  {journal} {Physical Review B}\ }\textbf
		{\bibinfo {volume} {52}},\ \bibinfo {pages} {6301} (\bibinfo {year}
		{1995})}\BibitemShut {NoStop}%
\end{thebibliography}
%
\end{document}